\documentclass[11pt]{article}
\usepackage{graphicx, amssymb}%, showkeys}
\usepackage[mathscr]{eucal}
\usepackage{array,amsmath,amsfonts,epsfig,bbm,longtable,float}
\newcommand{\SetFigFont}[3]{}

\title{Fermion Systems in Discrete Space-Time Exemplifying
the Spontaneous Generation of a Causal Structure}

\author{A.\ Diethert, F.\ Finster and D.\ Schiefeneder}
\date{October 2007}

\newtheorem{Def}{Definition}[section]
\newtheorem{Thm}[Def]{Theorem}
\newtheorem{Prp}[Def]{Proposition}
\newtheorem{Lemma}[Def]{Lemma}

\newtheorem{Example}[Def]{Example}
\newcommand{\Proof}{{\em{Proof.}}}
\newcommand{\QED}{\ \hfill $\FBox$ \\[1em]}
\newcommand{\QEDrem}{\ \hfill $\blacklozenge$}
\newcommand{\spc}{\;\;\;\;\;\;\;\;\;\;}

\newcommand{\bra}{\mbox{$< \!\!$ \nolinebreak}}
\newcommand{\ket}{\mbox{\nolinebreak $>$}}
\newcommand{\C}{\mathbb{C}}
\newcommand{\R}{\mathbb{R}}
\newcommand{\1}{\mbox{\rm 1 \hspace{-1.05 em} 1}}
\newcommand{\Z}{\mathbb{Z}}

\newcommand{\Tr}{\mbox{\rm{Tr}\/}}

\newcommand{\beq}{\begin{equation}}
\newcommand{\eeq}{\end{equation}}
\newcommand{\FBox}{\rule{2mm}{2.25mm}}

\newcommand{\sgn}{{\mbox{\rm{sgn}}}}

\newcommand{\Pp}{{\mathcal{P}}}
\newcommand{\Ss}{{\mathcal{S}}}
\newcommand{\Ll}{{\mathcal{L}}}
\newcommand{\G}{{\mathcal{G}}}
\newcommand{\Oo}{{\mathcal{O}}}
\newcommand{\Zz}{{\mathcal{Z}}}

\begin{document}
\maketitle
\begin{abstract}
As toy models for space-time on the Planck scale,
we consider examples of fermion systems in discrete space-time
which are composed of one or two particles defined on two up to nine
space-time points.
We study the self-organization of the particles as described by a variational principle
both analytically and numerically.
We find an effect of spontaneous symmetry breaking which leads to the
emergence of a discrete causal structure.
\end{abstract}
\tableofcontents

\newpage
\section{Introduction}
\setcounter{equation}{0}  \label{sec1}
Following ideas from~\cite{PFP}, fermion systems in discrete space-time were introduced
in~\cite{F1} as systems of a finite number of quantum mechanical particles defined
on a finite number of space-time points (see also the review articles~\cite{F2, F5}).
The interaction of the particles is described by a variational principle, where we minimize
an action defined on the ensemble of the corresponding wave functions.
A-priori, there are no relations between the space-time points; in particular, there is
no nearest-neighbor relation and no notion of causality. The idea is that these 
additional structures should be generated spontaneously.
More precisely, in order to minimize the action, the wave functions form specific
configurations; this can be visualized as a ``self-organization'' of the particles.
As a consequence of this self-organization,
the wave functions induce non-trivial relations between the
space-time points. The hope is that these relations give rise to additional structures,
which, in a suitable limit where the number of particles and space-time points
tends to infinity, can be identified with the local and causal structure of Minkowski space
or a Lorentzian manifold. In this limit, the configuration of the wave functions should go over
to a Dirac sea structure.

This intuitive picture has been made precise to some extent in the following papers.
In~\cite{F1} the variational principle in discrete space-time is analyzed, and it is shown
that minimizers exist. In~\cite{F3} the symmetries of discrete fermion systems
are studied systematically. It is proved under general assumptions, which include all
cases of physical interest, that the permutation symmetry of the space-time points cannot
be respected by the wave functions. In other words, the permutation symmetry of discrete
space-time is spontaneously broken by the particles. This shows that the wave functions
indeed induce non-trivial relations between the space-time points. The paper~\cite{F4} is
devoted to the transition from discrete space-time to Minkowski space. The method is to
analyze regularizations of vacuum Dirac seas in composite expressions which arise in our
variational principle. Finally, in~\cite{PFP} we study the so-called continuum limit.
Simply assuming that the wave functions form a Dirac sea configuration, we get an effective
continuum theory where Dirac particles interact with classical gauge fields.
Our variational principle determines the corresponding gauge groups and the coupling
of the gauge fields to the Dirac particles. We thus obtain many promising concrete results
which seem to indicate that our variational principle really is of physical
significance.

The motivation for the present paper is that the representation theory
in~\cite{F3} does not give an intuitive understanding of how the mechanism of
spontaneous symmetry breaking works. Also, the abstract constructions do not
give information on what the resulting smaller symmetry group is.
More specifically, we would like to know whether we get relations between the space-time
points which can be interpreted as a discrete version of a causal structure.
In order to get a better understanding of these important issues, we here
study concrete examples involving one or two particles.
Clearly, these systems are much too small for being of direct physical interest. But they
can be considered as interesting toy models, where the general mechanism of
spontaneous symmetry breaking and the generation of a discrete causal structure can
be studied in detail.

This paper begins with the basics and can be used as an easily accessible introduction
to fermion systems in discrete space-time. Whenever possible, we give explicit calculations
and self-consistent proofs. But we also present numerical results and refer to general
theorems in~\cite{F1, F3}, which instead of reproving we illustrate or explain
in words. 

\section{Preliminaries and Basic Definitions}
\setcounter{equation}{0}
In this section we give the basic definitions and collect a few general properties
of fermion systems in discrete space-time. For simplicity, we always restrict
attention to the case of spin dimension one, because all our examples will be constructed
in this case. For the generalization to higher spin dimension we refer to~\cite{F1, F3}.
Let~$H$ be a finite-dimensional complex vector space,
endowed with a sesquilinear form $\bra .|. \ket \::\:
H \times H \to \C$, i.e.\ for all $u, v, w \in H$ and $\alpha, \beta \in \C$,
\begin{eqnarray*}
\bra u \:|\: \alpha v + \beta w \ket &=& \alpha \:\bra u \:|\: v \ket +
\beta \:\bra u \:|\: w \ket \\
\bra \alpha u + \beta v \:|\: w \ket &=& \overline{\alpha} \:\bra u \:|\: w \ket
+ \overline{\beta} \:\bra v \:|\: w \ket\:.
\end{eqnarray*}
We assume that~$\bra .|. \ket$ is Hermitian symmetric,
\[ \overline{\bra u \:|\: v \ket} \;=\;\: \bra v \:|\: u \ket \:, \]
and non-degenerate,
\[ \bra u \:|\: v \ket \;=\; 0 \;\;\; \forall \:v \in H \quad \Longrightarrow \quad
u \;=\; 0 \:. \]
We also refer to~$(H, \bra .|. \ket)$ as an {\em{indefinite inner product space}}.
We point out that the inner product does not need to be positive, but it can have a general
signature~$(p,q)$ (see~\cite{Bognar, GLR} for details).
Nevertheless, many constructions from Hilbert spaces can be carried over.
In particular, we define the {\em{adjoint}} of
a linear operator~$A \::\: H \to H$ by the relation
\[ \bra u \:|\: A v \ket \;=\; \bra A^* u \:|\: v \ket \qquad
\forall \:u, v \in H\:. \]
A linear operator~$A$ is said to be {\em{unitary}} if~$A^*=A^{-1}$
and {\em{self-adjoint}} if $A^*=A$. It is called a {\em{projector}} if it is
self-adjoint and idempotent.

Let~$M$ be a finite set consisting of $m=\# M$ points. To every point~$x \in M$ we associate a projector
$E_x$. We assume that these projectors are orthogonal and
complete in the sense that
\beq \label{oc}
E_x\:E_y \;=\; \delta_{xy}\:E_x \spc {\mbox{and}} \spc
\sum_{x \in M} E_x \;=\; \1\:.
\eeq
Furthermore, we assume that the images~$E_x(H) \subset H$ of these
projectors are non-de\-ge\-ne\-rate subspaces of~$H$, which
all have the same dimension two and the same signature~$(1,1)$
(we remark that in the generalization to higher spin dimension, these subspaces have
signature~$(n,n)$ with~$n>1$). The points~$x \in M$ are
called {\em{discrete space-time points}}, and the corresponding
projectors~$E_x$ are the {\em{space-time projectors}}. The
structure~$(H, \bra .|. \ket, (E_x)_{x \in M})$ is
called {\em{discrete space-time}}.

To describe the particles of our system, we introduce one more projector~$P$ on~$H$,
the so-called {\em{fermionic projector}}, which has the additional property that its
image~$P(H)$ is {\em{negative}} definite.
The vectors in the image of~$P$ can be interpreted as the
quantum mechanical states of the particles of our system, and 
thus we call the rank of~$P$ the {\em{number of particles}}, and denote
this number by~$f=\dim P(H)$.
The name ``fermionic projector'' is motivated from the fact that in
physical applications~\cite{PFP}, the particles are Dirac particles and thus fermions.
We refer to~$(H, \bra .|. \ket, (E_x)_{x \in M}, P)$ as a {\em{fermion system
in discrete space-time}}.

A space-time projector $E_x$ can be used to restrict an operator to the subspace $E_x(H)$.
We refer to this restriction as the \textit{localization} at the space-time point $x$. 
Localizing the fermionic projector gives rise to the so-called \textit{discrete kernel}
$P(x,y):=E_x P E_y$, which maps the subspace $E_y(H) \subset H$ to $E_x(H)$ and vanishes otherwise. It is most convenient to regard it as a mapping only between these subspaces,
\[ P(x,y)\;:\; E_y(H) \: \rightarrow\: E_x(H)\:. \]
Then the  product of $P(x,y)$ and $P(y,x)$ is an endomorphism of $E_x(H)$,
\[ A_{xy}=P(x,y)P(y,x) \;:\; E_x(H) \:\rightarrow\: E_x(H) \:, \]
referred to as a \textit{closed chain}. The closed chain is a self-adjoint operator on~$E_x(H)$.
The roots of its characteristic polynomial counted with multiplicities are denoted
by~$\lambda_+$ and~$\lambda_-$. We define the \textit{spectral weight} of $A_{xy}$ by
\[ |A_{xy}| \;=\; |\lambda_+| + |\lambda_-| . \]
Similarly, the spectral weight of the square of the closed chain
is defined by~$|A_{xy}^2|= |\lambda_+|^2 + |\lambda_-|^2$.
Summing the spectral weights over all space-time points we get a positive number, which only depends on the fermionic projector relative to the space-time projectors.
For a given $\kappa > 0$ we consider the family of fermionic projectors
\beq \label{cfamily}
\Pp (\kappa):= \left\lbrace P \; \mbox{with} \sum_{x,y\in M} \vert A_{xy} \vert^2 = \kappa \right\rbrace
\eeq
and introduce the function
\beq \label{target}
\Zz[P] \;=\; \sum_{x,y\in M} \vert A_{xy}^2 \vert \:.
\eeq
Our variational principle is to
\begin{equation}\label{varpr}
\mbox{minimize $\Zz[P]$ by varying} \; P \; \mbox{in} \; \Pp(\kappa)\:,
\end{equation}
keeping the number of particles~$f$ as well as discrete space-time fixed.
For clarity, we sometimes refer to~(\ref{varpr}) as the {\em{variational principle
with constraint}}.

Using the method of Lagrange multipliers, for every minimizer $P$ there is a real parameter $\mu$ such that $P$ is a stationary point of the so-called \textit{action}
\beq \label{Smdef}
\Ss_{\mu}[P] \;=\; \sum_{x,y\in M} \Ll_{\mu} [A_{xy}]
\eeq
with the \textit{Lagrangian}
\beq \label{Lagdef}
\Ll_{\mu}[A_{xy}] \;:=\; \vert A_{xy}^2 \vert - \mu\; \vert A_{xy}\vert ^2 \:.
\eeq
A possible method for constructing stationary points is to minimize~$\Ss_{\mu}$, and this
leads us to the so-called \textit{auxiliary variational principle}
\begin{equation}\label{auxvp}
	\mbox{ minimize }\; \Ss_{\mu}[P] \; \mbox{ by varying }\; P\:.
\end{equation}

Without particles~($f=0$), the variational principles~(\ref{varpr}) and~(\ref{auxvp}) are
of course trivial. If on the other hand there is only one space-time point ($m=1$),
there is only one closed chain~$A=P^2=P$ and hence
\[ |A^2| \;=\; |A| \;=\; f\:, \qquad S_\mu \;=\; f - \mu \,f^2 \:. \]
Thus the variational principles~(\ref{varpr}) and~(\ref{auxvp}) are again trivial.
Therefore, we shall always assume that
\[ f \;\geq\; 1\qquad {\mbox{and}} \qquad m \;>\; 1\:. \]

Before discussing the above variational principles, 
it is convenient to bring the fermion system into a more explicit form by writing down the
projectors~$(E_x)_{x \in M}$ and~$P$ as matrices in a given basis of~$H$.
We shall always work with a basis of the following form:
In each subspace~$E_x(H) \subset H$ we choose a pseudo-orthonormal
basis~$(e^x_1, e^x_2)$, i.e.
\[ \bra e^x_i \,|\, e^x_j \ket \;=\; s_i\, \delta_{ij} \qquad {\mbox{with}} \qquad
s_1=1\:, s_2=-1\:. \]
The orthogonality relation in~(\ref{oc}) yields that $\bra e^x_i| e^y_j \ket=0$
for all~$x \neq y$ and all~$i,j \in \{1,2\}$. Furthermore, the completeness
relation in~(\ref{oc}) yields that the vectors~$(e^x_i)^{x \in M}_{i \in \{1,2\}}$
form a basis of~$H$. In particular, we see that~$H$ has dimension $2m$ and
signature~$(m, m)$.
Ordering the basis vectors as~$(e^1_1, e^1_2, e^2_1, e^2_2, \ldots)$, the inner product
of~$H$ has the representation
\beq \label{iprep}
\bra u \,|\, v \ket \;=\; (u \,|\, S\, v) \qquad \forall \,u, v \in H \:,
\eeq
where the round brackets denote the standard Euclidean scalar product on~$\C^{2m}$,
and the so-called {\em{signature matrix}} $S$ in block matrix notation is given by
\beq \label{basis1}
S \;=\; \begin{pmatrix}
	s & &   \\[-.5em]
	 & \ddots &  \\[-.5em]
	 & & s
\end{pmatrix} \qquad \mbox{with} \qquad s \;=\; \begin{pmatrix}
	1 & 0 \\
	0 & -1
	\end{pmatrix} \:.
\eeq
In the same block matrix notation, the space-time projectors become
\beq \label{basis2}
E_1=\left( \begin{smallmatrix}
	\mathbbm{1} & & &  \\
	 & 0 & & \\[-0.5em]
	 & & \ddots & \\
	 & & &0
\end{smallmatrix} \right), \quad
E_2=\left( \begin{smallmatrix}
		0 & & & & \\
		& \mathbbm{1} & &  \\
		& & 0 & & \\[-0.5em]
		& & & \ddots & \\
		& & & & 0
\end{smallmatrix} \right),  \ldots , \qquad
E_m=\left(  \begin{smallmatrix}
	0 & & &  \\[-0.5em]
	 & \ddots & & \\
	 & & 0 & \\
	 & & & \mathbbm{1}
\end{smallmatrix} \right) .
\eeq
It is instructive to briefly discuss the fermionic projector and the closed chain in this basis.
Since~$P(H)$ is negative definite, the number of particles
must be bounded by the negative signature of~$H$, i.e.\ we always have~$f \leq m $.
The self-adjointness of~$P$ can be expressed by
\beq \label{Psymm}
(PS)^\dagger \;=\; PS\:,
\eeq
where the dagger denotes the transpose, complex conjugate matrix.
For any~$x, y \in M$, the closed chain~$A_{xy}$ is a self-adjoint operator on~$E_x(H)$.
Thus it is expressed by a $(2 \times 2)$-matrix satisfying the relation
\[ (A_{xy}\, s)^\dagger \;=\; A_{xy}\, s\:. \]
Applying the identity~$\det(B C - \lambda \1) = \det(C B - \lambda \1)$ to
the characteristic polynomial of~$A_{xy}$, we find that
\begin{eqnarray*}
\det(A_{xy} - \lambda \1) &=& \det(P(x,y)\, P(y,x) - \lambda \1) \\
&=& \det(P(y,x)\, P(x,y) - \lambda \1) \;=\; \det(A_{yx} - \lambda \1) \:,
\end{eqnarray*}
proving that the operators~$A_{xy}$ and~$A_{yx}$ have the same
characteristic polynomial. This shows that our Lagrangian is symmetric in
the two space-time points,
\beq \label{symmetry}
{\mathcal{L}}_\mu[A_{xy}] \;=\; {\mathcal{L}}_\mu[A_{yx}] \spc
\forall \, x,y \in M\:.
\eeq
In contrast to Hilbert spaces, the self-adjointness of a closed chain~$A$ does not imply
that its eigenvalues are real. But the calculation
\[ \overline{ \det(A-\lambda\1) } \;=\;
\det \!\left( A^\dagger - \overline{\lambda} \right) \;=\;
\det \!\left( s (A^\dagger - \overline{\lambda})s \right) \;=\;
\det \!\left( A^* - \overline{\lambda} \right) \;=\;
\det \!\left( A - \overline{\lambda}\1 \right) \]
shows that its characteristic polynomial has real coefficients. This means that
the~$\lambda_\pm$ are either both real, or else they form a complex conjugate pair.

We now give a brief overview of the existence theory.
In~\cite{F1} it is proved that the variational principle with constraint~(\ref{varpr}) is well-posed.
\begin{Thm}
If $\mathcal{P}(\kappa)$ is non-empty, the variational principle (\ref{varpr}) attains its minimum.
\end{Thm}
The auxiliary variational principle behaves differently depending on the value of~$\mu$,
as the following consideration shows.
\begin{Prp}
If~$\mu \leq \frac{1}{2}$, the Lagrangian~$\Ll_\mu$ is non-negative.
If conversely~$\mu > \frac{1}{2}$ and~$f \geq 2$, the action~$\Ss_\mu$ is not bounded
from below.
\end{Prp}
{\Proof} 
The Cauchy-Schwarz inequality~$2 |\lambda_+ \lambda_-| \leq |\lambda_+|^2+|\lambda_-|^2$
yields
\[ |A|^2 \;=\; \left( |\lambda_+| + |\lambda_-| \right)^2 \;=\;
|\lambda_+|^2 + 2 |\lambda_+ \lambda_-| + |\lambda_-|^2 \;\leq\;
2 |\lambda_+|^2 + 2 |\lambda_-|^2 \;=\; 2\, |A^2|\:.\]
Hence
\[ \Ll_\mu \;=\; |A^2| - \mu\, |A|^2 \;\geq\; \left( \frac{1}{2} - \mu \right) |A|^2\:, \]
proving that the Lagrangian is non-negative in the case~$\mu \leq \frac{1}{2}$.

In the remaining case~$\mu > \frac{1}{2}$, we shall construct a family of
fermionic projectors where~$\Ss_\mu$ tends to minus infinity.
In preparation, we briefly consider the simple case when~$H$ has signature~$(1,1)$.
In this case, the fermionic projector must be of rank one. It is most conveniently
represented by choosing a vector~$u \in H$ with~$\bra u \,|\, u \ket = -1$ and to
set
\[ P \;=\; - |u \ket \bra u| \;=\; - u \otimes u^\dagger S\:, \]
where we used a bra/ket-notation and then represented the inner product in the form~(\ref{iprep})
with~$S={\mbox{diag}}(1,-1)$. Choosing~$u=(\sqrt{\alpha}, \sqrt{\alpha+1})^T$ with a parameter~$\alpha>0$, we
obtain the fermionic projector with the matrix representation
\beq \label{Psimple}
P \;=\; \left( \begin{matrix} -\alpha & \beta \\ -\beta & \alpha+1 \end{matrix} \right)
\quad {\mbox{ with~$\beta=\sqrt{\alpha(\alpha+1)}$}}.
\eeq

Coming back to the situation of the proposition, we know that~$f \geq 2$ and thus~$m \geq 2$.
Working in the matrix representation~(\ref{basis1},
\ref{basis2}), we choose a fermionic projector which is invariant on the
subspace~$E_1(H) \oplus E_2(H)$, and on this subspace has the form
\[ P |_{E_1(H) \oplus E_2(H)} \;=\;
\left( \begin{matrix}
	-\alpha & 0 & 0 & \beta  \\
	 0 & \alpha+1 & -\beta & 0 \\
	 0 & \beta & -\alpha & 0 \\
	 -\beta & 0 & 0 & \alpha+1
\end{matrix} \right) . \]
Since this matrix is built up of components of the form~(\ref{Psimple}), it is
obviously a projector on a two-dimensional, negative definite subspace. 
Thus, choosing~$P$ on the orthogonal complement~$(E_1(H) \oplus E_2(H))^\perp$ equal to
any fixed projector on a negative definite subspace of dimension~$f-2$, we obtain a fermionic
projector.

Computing the closed chains, we obtain
\[ A_{11}\;=\; A_{22} \;=\;  \alpha^2 \,\1 + {\mathcal{O}}(\alpha) \:,\qquad
A_{12} = A_{21} \;=\; -\alpha^2\, \1 + {\mathcal{O}}(\alpha)\:. \]
whereas all other closed chains are independent of~$\alpha$. Hence
\[ \sum_{x,y \in M} |A^2_{xy}| \;=\; 4\, \alpha^4\, |\1_{\C^2}| + {\mathcal{O}}(\alpha^3)\:, \qquad
\sum_{x,y \in M} |A_{xy}|^2 \;=\; 4\, \alpha^4\, |\1_{\C^2}|^2 + {\mathcal{O}}(\alpha^3)  \:. \]
Using that~$|\1_{\C^2}|=2$, we obtain for the action the asymptotic formula
\[ \Ss_\mu \;=\; 8\, \alpha^4\, (1-2 \mu) + {\mathcal{O}}(\alpha^3) \, , \]
and this really tends to minus infinity as~$\alpha \rightarrow \infty$.
\QED
According to this proposition, in the case~$\mu>\frac{1}{2}$
the auxiliary variational principle is not well-posed
(except if there is only one particle, see Proposition~\ref{prpone} below).
In the case~$\mu<\frac{1}{2}$ a general existence theorem is proved in~\cite{F1}.
\begin{Thm}
If $\mu< \frac{1}{2}$, the auxiliary variational principle attains its minimum.
\end{Thm}
In the remaining so-called {\em{critical case}}~$\mu=\frac{1}{2}$ of the auxiliary variational
principle, partial existence results are given in~\cite{F1}, but the general
existence problem is still open. From the physical point of view, the critical case is the
most interesting case, because it allows to model a system of massive Dirac seas (see~\cite{F4,
PFP}). For this reason, we introduce the abbreviation~$ \Ll \equiv \Ll_{\tfrac{1}{2}}$
and~$\Ss \equiv \Ss_{\tfrac{1}{2}}$. Using the transformation
\[ |A^2| - \frac{1}{2}\: |A|^2 \;=\;
\left( |\lambda_+|^2 + |\lambda_-|^2 \right) -\frac{1}{2} \left(  |\lambda_+| + |\lambda_-| \right)^2
\;=\; \frac{1}{2} \left(  |\lambda_+| -  |\lambda_-| \right)^2 \:, \]
the critical Lagrangian can be written in the simple form
\beq \label{Lsimp}
\Ll[A] \;=\; \frac{1}{2} \left(  |\lambda_+| -  |\lambda_-| \right)^2\:.
\eeq
Thus one can say that the critical case of the auxiliary variational principle tries to
achieve that the~$\lambda_\pm$ have the same absolute value.

We next discuss the symmetry structure of discrete fermion systems (for details see~\cite{F3}).
It is useful to distinguish between inner and outer symmetries. Inner symmetries are
also called {\em{gauge symmetries}} and keep~$M$ fixed, whereas
outer symmetries also involve a transformation of the space-time points.
A unitary mapping~$U$ of~$H$ is called a {\em{gauge transformation}} if it
leaves the space-time projectors unchanged, i.e.
\[ U E_x U^{-1} \;=\; E_x \qquad {\mbox{$\forall\, x \in M$}}. \]
The fermionic projector behaves under gauge transformations as follows,
\beq \label{Pgt}
P \;\longrightarrow\; U P U^{-1}\:.
\eeq
If~$P=UPU^{-1}$, the transformation~$U$ describes a gauge symmetry.
The group of all gauge transformations is denoted by~$\G$. 
In the basis~(\ref{basis1}, \ref{basis2}), a gauge transformation~$U$ can be written as
\beq \label{gt}
U \;=\; \begin{pmatrix} U_1 & & 0   \\ & \ddots &  \\ 0 &  & U_m
\end{pmatrix} \qquad \mbox{with} \qquad U_i \in U(1,1) \:.
\eeq
This means that a gauge transformation splits into the direct sum of  unitary transformations
which act "locally" on the spaces~$E_x(H)$. We also see that~$\G$ is isomorphic
to the group~$U(1,1)^m$.

To describe outer symmetries, we permute the space-time points and demand that,
after a suitable unitary transformation, the discrete fermion system should be unchanged.
We denote the symmetric group of~$M$ (i.e.\ the group of all permutations) by~$S_m$.
\begin{Def} \label{defouter}
A subgroup~$\Oo$ of the symmetric group~$S_m$
is called {\bf{outer symmetry group}} of the discrete fermion system if
for every~$\sigma \in \Oo$ there is a unitary
transformation~$U$ such that
\beq \label{USdef}
UPU^{-1} \;=\; P \qquad {\mbox{and}} \qquad
U E_x U^{-1} \;=\; E_{\sigma(x)} \quad
\forall\, x \in M\:.
\eeq
\end{Def}
We next specialize~\cite[Theorem~9.1]{F3} to the case of spin dimension one.
\begin{Thm} \label{sym}
Let~$(H, \bra .|. \ket, (E_x)_{x \in M}, P)$ be a fermion system in discrete
space-time, where the subspaces~$E_x(H) \subset H$ have signature~$(1,1)$.
Suppose that the number of particles~$f$ and the number
of space-time points~$m$ lie in the range
\[ 1 \;<\; f \;<\; m-1 \:. \]
Then the system cannot have the outer symmetry group~$S_m$.
\end{Thm}
This result has the interpretation that the permutation symmetry of
discrete space-time is spontaneously broken by the fermionic projector.
It implies that the fermionic projector induces non-trivial relations between the
space-time points. Unfortunately, the theorem does not give information on how these
additional structures look like. One would hope for a structure which
can be regarded as a discrete analogue of a causal structure.
Such a structure can indeed be defined in general. Recall that after~(\ref{symmetry})
we saw that the~$\lambda_\pm$ are either both real or else they form a complex
conjugate pair. This distinction can be used to introduce a notion of causality.
\begin{Def} {\bf{(discrete causal structure)}} \label{dcs}
Two discrete space-time points~$x,y \in M$ are called {\bf{timelike}}
separated if the roots~$\lambda_\pm$ of the characteristic polynomial
of~$A_{xy}$ are both real. Conversely, they are said to be
{\bf{spacelike}} separated if the~$\lambda_\pm$ form a complex
conjugate pair.
\end{Def}
This definition was first given in~\cite{F5}, and in~\cite{F2, F4} it is shown
that for Dirac spinors in Minkowski space it gives back the usual notion
of causality. The definition of the discrete causal structure can also be
understood from the fact that it reflects the structure of the critical Lagrangian~(\ref{Lsimp}).
Namely, if~$x$ and~$y$ are spacelike separated, the roots~$\lambda_\pm$
of the characteristic polynomial of~$A_{xy}$ form a complex conjugate
pair. Hence~$|\lambda_+| = |\overline{\lambda_-}| = |\lambda_-|$,
and thus the Lagrangian~(\ref{Lsimp}) vanishes.
Computing first variations of the Lagrangian, one sees that
these also vanish, and thus~$P(x,y)$ does not enter the Euler-Lagrange equations.
This can be seen in analogy to the usual notion of causality
in Minkowski space where points with spacelike separation 
cannot influence each other.

The next simple proposition illustrates our definition of a discrete causal structure.
The method is worked out in a more general context in~\cite[Section~4]{F1}
and is used in the existence proof.
\begin{Prp} Every space-time point has timelike separation from itself.
\end{Prp}
{\Proof} For any~$x \in M$, we can use the idempotence and self-adjointness of our
projectors to write the expectation value of~$P(x,x)$ as follows,
\[ \bra u \,|\, P(x,x)\, u \ket \;=\; \bra u \,|\, E_x P E_x\, u \ket
\;=\; \bra u \,|\, E_x P^2 E_x\, u \ket \;=\; \bra P E_x u \,|\, P E_x\, u \ket\:. \]
Since the image of~$P$ is negative definite, it follows that
\[ \bra u \,|\, P(x,x)\, u \ket \;\leq\; 0 \qquad \forall\, u \in H\:. \]
Expressed in the basis~(\ref{basis1}, \ref{basis2}), we conclude that the matrix
$-s P(x,x)$ is positive semi-definite on the standard Euclidean~$\C^2$.
This implies that its determinant is non-negative. Using that the determinant is
multiplicative and that~$\det (-\1_{\C^2})=1$ and~$\det s=-1$, we conclude that
\beq \label{detn}
\det P(x,x) \;\leq\; 0 \:.
\eeq
The operator~$P(x,x)$ is self-adjoint on~$E_x(H) \subset H$. Following the consideration
after~(\ref{symmetry}), the roots~$\mu_\pm$ of its characteristic polynomial
either are real or they form a complex conjugate pair. In the latter case, the determinant
of~$P(x,x)$ would be strictly positive, in contradiction to~(\ref{detn}). We conclude
that the~$\mu_\pm$ are real. Since
\[ A_{xx} \;=\; P(x,x)^2\:, \]
the spectral calculus yields that~$\lambda_\pm = \mu_\pm^2 \in \R$.
\QED

The aim of the present paper is to specify the spontaneous symmetry breaking of
Theorem~\ref{sym} and to clarify the connection to the discrete causal structure of
Definition~\ref{dcs} in simple examples.

\section{One-Particle Systems}
\setcounter{equation}{0}
In this section we will show that for systems of only one particle, our variational principles
can be solved analytically. We first remark that, since~$P$ has rank one,
for any~$x, y \in M$ the discrete kernel~$P(x,y)$ has rank at most one.
Thus the closed chains~$A_{xy}$ also have rank at most one, and thus
\beq \label{Asid}
|A_{xy}^2| \;=\; |A_{xy}|^2 \;=\; \Tr(A_{xy})^2 \qquad \forall\, x,y \in M\:.
\eeq
This implies that the functional in the variational principle with constraint~(\ref{varpr})
is equal to~$\kappa$, making the variational principle trivial.
\begin{Prp} \label{prptrivial}
In the case~$f=1$ of one particle, every~$P \in {\mathcal{P}}(\kappa)$ is a minimizer
of the variational principle~(\ref{varpr}). 
\end{Prp}
Using~(\ref{Asid}), the Lagrangian of the auxiliary variational principle~(\ref{Lagdef})
simplifies to
\[ \Ll_\mu[A_{xy}] \;=\; (1-\mu)\, \Tr(A_{xy})^2\:.  \]
To further simplify the action, we introduce for any~$x \in M$
the {\em{local trace}} $\rho_x$ by
\beq \label{rhodef}
\rho_x \;=\; \Tr(E_x P) \in \R \:.
\eeq
Next we choose a vector~$u \neq 0$ in the image of~$P$ and
normalize it to~$\bra u \,|\, u \ket =-1$. Then the fermionic projector can be written in
bra/ket-notation as
\beq \label{Pbk}
P \;=\; -|u \ket \bra u|\:. 
\eeq
Furthermore, the local trace becomes~$\rho_x = -\bra u | E_x | u \ket$, and thus
\[ \Tr(A_{xy}) \;=\; \Tr(E_x \, P\, E_y\, P) \;=\;
\bra u | E_x |u \ket \,\bra u| E_y |u \ket \;=\; \rho_x\, \rho_y\:. \]
This simple factorization allows us to write the whole action as a square,
\beq \label{Smform}
{\mathcal{S}}_\mu \;=\; (1-\mu)\, \sum_{x,y \in M} \Tr(A_{xy})^2
\;=\; (1-\mu)\, \sum_{x,y \in M} \rho_x^2 \rho_y^2 \;=\;
 (1-\mu) \left( \sum_{x \in M} \rho_x^2 \right)^2\:.
 \eeq
 Now we can use the Cauchy-Schwarz inequality to obtain the following result.
 \begin{Prp} \label{prpone} Consider the auxiliary variational principle~(\ref{auxvp}) for
 fermion systems in discrete space-time with one particle. 
 Depending on the value of~$\mu$, there are the following cases:
 \begin{description}
 \item[(i)] If~$\mu=1$, $\Ss_\mu$ vanishes identically, and
 every fermionic projector is a minimizer.
 \item[(ii)] If~$\mu<1$, the minimum is attained. We can arrange by a gauge transformation
 that in the basis~(\ref{basis1}, \ref{basis2}), the minimizing fermionic projector  can be written
 as~(\ref{Pbk}) with
 \beq \label{Exucond}
 E_x\, u \;=\; \frac{1}{\sqrt{m}} \left( \begin{matrix} 0 \\ 1 \end{matrix} \right)
 \qquad \forall\, x \in M \:.
 \eeq
 \item[(iii)] If~$\mu>1$ and~$m>1$, the action is not bounded from below.
\end{description}
 \end{Prp}
{\Proof} The case~{\bf{(i)}} is obvious from~(\ref{Smform}). To prove~{\bf{(iii)}},
we consider for~$\alpha>0$ the fermionic projector~(\ref{Pbk}) with
\[ E_1\, u \;=\; \left( \begin{matrix} \sqrt{\alpha} \\ 0 \end{matrix} \right) , \qquad
E_2\, u \;=\; \left( \begin{matrix} 0 \\ \sqrt{\alpha+1} \end{matrix} \right) , \qquad
E_y\, u \;=\; 0 \quad \forall y\, \in \{3, \ldots, m\} . \]
A short calculation shows that~$\Ss_\mu$ as given by~(\ref{Smform})
tends to minus infinity as~$\alpha \rightarrow \infty$.

In the remaining case~{\bf{(ii)}}, we apply the completeness relation in~(\ref{oc})
to~(\ref{rhodef}), giving
\[ \sum_{x \in M} \rho_x \;=\; \Tr\, P \;=\; 1\:. \]
Hence we can apply the Cauchy-Schwarz inequality to obtain
\beq \label{CS}
1 \;=\; \left(\sum_{x \in M} \rho_x \right)^2 \;\leq\; \left(\sum_{x \in M} 1 \right)
\left(\sum_{x \in M} \rho_x^2 \right) \;=\; m \sum_{x \in M} \rho_x^2\:.
\eeq
Using this inequality in~(\ref{Smform}) gives the lower bound
\[ \Ss_\mu \;\geq\; \frac{1-\mu}{m^2}\:, \]
and equality holds if and only if the~$\rho_x$ are all equal. Hence every minimizer must satisfy
the relations
\[ \rho_x \;=\; \frac{1}{m}\quad \forall\, x \in M \:. \]
This means that the components of the vector $E_x u=:(a,b)^T$ satisfy the
relation~$a^2-b^2=-1/m$. Thus by an unitary transformation on~$E_x(H)$ we can
arrange~(\ref{Exucond}).
\QED
We remark that this proposition immediately generalizes to higher spin dimension.

The minimizer in case~{\bf{(ii)}} has the property that the particle is
completely delocalized in space-time, because according to~(\ref{Exucond})
it has the same probability~$1/m$ to be at any of the space-time points.
The resulting system is obviously permutation symmetric, and thus no spontaneous symmetry
breaking occurs. Furthermore, as the closed chain~$A_{xy}$ has rank at most one,
one of the roots of its characteristic polynomial must vanish. Hence the~$\lambda_\pm$
cannot form a complex conjugate pair. Following Definition~\ref{dcs}, all
pairs of space-time points are timelike separated, and the discrete causal
structure is trivial.

\section{Two-Particle Systems in the Critical Case}
\setcounter{equation}{0}
\subsection{The Local Correlation Matrices and the Fermion Matrix}
We begin the discussion of many-particle systems with the simplest situation
of two particles ($f=2$) and the auxiliary variational principle in the
critical case~(\ref{auxvp}, \ref{Smdef}, \ref{Lsimp}).
For systems of two particles, the following construction is very useful
for visualizing the discrete fermion system.
The image of~$P$ is a two-dimensional, negative definite subspace of~$H$.
Choosing an orthonormal basis $(u_1,u_2)$ (i.e.\ $\bra u_i | u_j \ket = -\delta_{ij}$),
the fermionic projector can be written in bra/ket-notation as
\beq \label{2braket}
P \;=\; -|u_1 \ket \bra u_1| - |u_2 \ket \bra u_2|\:.
\eeq
For any space-time point~$x \in M$, we can introduce the
so-called {\em{local correlation matrix}} $F_x$ by
\beq \label{Fxdef}
F_x \;=\; ((F_x)^i_j)_{i,j=1,2} \quad {\mbox{with}} \quad
(F_x)^i_j \;=\; -\bra u_i \,|\, E_x u_j \ket \:.
\eeq
The matrix~$F_x$ is Hermitian on the standard Euclidean $\C^2$. Thus we
can decompose it in the form
\beq \label{Fxd}
F_x \;=\; \frac{1}{2} \left( \rho_x\, \1 + \vec{v}_x \vec{\sigma} \right) ,
\eeq
where~$\vec{\sigma}=(\sigma^1, \sigma^2, \sigma^3)$ are the Pauli matrices,
and~$\rho_x \in \R$, $\vec{v}_x \in \R^3$. Taking the trace of~(\ref{Fxd})
and using~(\ref{Fxdef}), we find
\[ \rho_x \;=\; -\sum_{i=1,2} \bra u_i \,|\, E_x u_i \ket \;=\;
\Tr(E_x P)\:, \]
showing that~$\rho_x$ is again the local trace~(\ref{rhodef}).
The vector~$\vec{v}_x$ describes the correlations of the two particles at the
space-time point~$x$. Using the terminology which is common in atomic physics
and in quantum information theory for the description of q-bits,
we refer to~$\vec{v}_x$ as the {\em{Bloch vector}}.
Summing~(\ref{Fxdef}) over~$x \in M$ and using the completeness relation in~(\ref{oc}),
we find that~$\sum_{x \in M} F_x =\1$ or, equivalently,
\beq \label{rhovcond}
\sum_{x \in M} \rho_x \;=\; 2 \qquad {\mbox{and}} \qquad \sum_{x \in M} \vec{v}_x
\;=\; \vec{0}\:.
\eeq
Furthermore, as the inner product in~(\ref{Fxdef}) has signature~$(1,1)$, the
matrix~$F_x$ can have at most one positive and at most one negative eigenvalue.
Expressed in terms of the decomposition~(\ref{Fxd}), this means that
\beq \label{rhovc2}
|\vec{v}_x| \;\geq\; \rho_x \qquad \forall\, x \in M\:.
\eeq
We point out that, since in~(\ref{Fxdef}) we took expectation values, the
local trace and the Bloch vector are gauge invariant quantities. However, our
construction clearly involves the arbitrariness of choosing the orthonormal
basis~$(u_1, u_2)$ of~$P(H)$. More precisely, we have the freedom to
unitarily transform the basis as follows,
\beq \label{U2}
u_i \;\longrightarrow\; \sum_{j=1}^2 V_{ij} u_j \qquad {\mbox{with}} \qquad
V \in U(2)\:.
\eeq
As is well-known from the transformation law for Pauli spinors in non-relativistic quantum
mechanics (see for example~\cite[Section 3.2]{QMbook}),
a unitary transformation of Pauli matrices corresponds to an $SO(3)$-transformation of the vector index~$\alpha \in \{1,2,3\}$,
\beq \label{Rot}
V \sigma^\alpha V^{-1} \;=\; \sum_{\beta=1}^3 R^{\alpha \beta} \,\sigma^\beta
\qquad {\mbox{with}} \qquad R \in SO(3)\:,
\eeq
and thus the $U(2)$-transformation (\ref{U2}) leads to a joint rotation of all Bloch vectors,$$v_x^{\alpha}\longrightarrow \sum_{\beta=1}^3 v_x^{\beta} R^{\beta \alpha} \qquad \forall \,x\in M.$$
Hence we can say that the Bloch vectors~$\vec{v}_x \in \R^3$
are unique up to rotations (i.e.\ orientation-preserving isometries) of
the standard Euclidean~$\R^3$. 

For calculations, it is convenient to choose the basis~(\ref{basis1}, \ref{basis2})
and to introduce a matrix whose columns are the components of
the orthonormal basis vectors~$u_1$ and~$u_2$,
\beq \label{Psidef}
\Psi \;:=\; \left( \begin{matrix} (u_1)_1 & (u_2)_1 \\
(u_1)_2 & (u_2)_2 \\
\vdots & \vdots \\
(u_1)_{2m} & (u_2)_{2m} \end{matrix} \right) .
\eeq
The matrix~$\Psi$ is referred to as the {\em{fermion matrix}}.
Clearly, the fermion matrix characterizes the fermion system completely.
Gauge transformations~(\ref{gt}) act on~$\Psi$ from the left, whereas the
$U(2)$-transformations~(\ref{U2}) act on it from the right. Thus we have the
freedom to transform the fermion matrix according to
\beq \label{Psif}
\Psi \;\longrightarrow\; U \,\Psi\, V^{-1} \qquad {\mbox{with $U \in \G$
and~$V \in U(2)$}}\,.
\eeq
Using the fermion matrix, we can write~(\ref{2braket}) in the compact form
\beq \label{PsiP}
P \;=\; -\Psi \Psi^\dagger S \:.
\eeq
For any~$x \in M$, the local correlation matrix becomes
\beq \label{PsiFx}
F_x \;=\; -\Psi^\dagger S E_x \Psi\:,
\eeq
whereas the local trace and the Bloch vectors can be computed by
\beq \label{Psirv}
\rho_x \;=\; -\Tr (\Psi^\dagger S E_x \Psi)\:,\qquad
v_x^\alpha \;=\; -\Tr (\sigma^\alpha\, \Psi^\dagger S E_x \Psi)\:.
\eeq
Note that when applying the transformation~(\ref{Psif}) in~(\ref{PsiP}),
the matrix~$V$ drops out, leaving us with the gauge freedom~(\ref{Pgt}).
In~(\ref{PsiFx}), on the other hand, the matrix~$U$ drops out, so that
we are left with the $U(2)$-freedom~(\ref{U2}).

We now collect a few results which clarify the significance of the
local correlation matrices~$F_x$. First, it is a useful fact that our
variational principles can be expressed in terms of the local correlation matrices.
\begin{Prp} \label{prp41} The roots~$\lambda_\pm$ of the characteristic polynomial of the
closed chain~$A_{xy}$ are written in terms of the local traces and the Bloch vectors by
\beq \label{lpm}
\lambda_\pm \;=\; \frac{1}{4} \, \left( \rho_x \rho_y + \vec{v}_x \vec{v}_y \:\pm\:
\sqrt{|\rho_x \vec{v}_y + \rho_y \vec{v}_x|^2 - |\vec{v}_x \times \vec{v}_y|^2 } \right) .
\eeq
Furthermore,
\beq \label{ssign}
\lambda_+ \lambda_- \;\geq\; 0 \:.
\eeq
\end{Prp}
{\Proof} Using~(\ref{PsiP}), the closed chain can be written as the operator product
\[ A_{xy} \;=\; E_x \, \Psi \Psi^\dagger S\, E_y\, \Psi \Psi^\dagger S\, E_x \:. \]
Using the identity~$\det(B C - \lambda \1) = \det(C B - \lambda \1)$, 
a cyclic commutation of the factors in the operator product does not change the spectrum.
Hence the $\lambda_\pm$ are also the roots of the characteristic polynomial of the matrix
\[ \Psi^\dagger S \,E_x \Psi \; \Psi^\dagger S \,E_y \Psi \;\stackrel{(\ref{PsiFx})}{=}\; F_x F_y\:. \]

Using the decomposition~(\ref{Fxd}) of the local correlation matrix together with the
identities of Pauli matrices (see~\cite[Section 3.2]{QMbook})
\[ \sigma^\alpha \sigma^\beta \;=\; \delta^{\alpha \beta} + \sum_{\gamma=1}^3
i \epsilon^{\alpha \beta \gamma} \sigma^\gamma \]
(where~$\epsilon^{\alpha \beta \gamma}$ is the totally anti-symmetric Levi-Civita symbol),
we obtain
\begin{eqnarray*}
F_x F_y &=& \frac{1}{4} \left(\rho_x + \vec{v}_x \vec{\sigma}  \right)
\left(\rho_y+ \vec{v}_y \vec{\sigma}  \right) \\
&=&  \frac{1}{4} \, \Big( \rho_x \rho_y + \vec{v}_x \vec{v}_y
\:+\: \left(\rho_x \vec{v}_y + \rho_y \vec{v}_x + i \vec{v}_x \times \vec{v}_y \right)
\vec{\sigma} \Big)
\end{eqnarray*}
(where the cross product is defined by $(\vec{x} \times \vec{y})^\gamma =
\epsilon^{\alpha \beta \gamma} x^\alpha y^\beta$).
Using that the vector~$\vec{v}_x \times \vec{v}_y$ is orthogonal to~$\vec{v}_x$ and~$\vec{v}_y$,
one easily verifies that
\[ \left[ F_x F_y - \frac{1}{4} \left( \rho_x \rho_y + \vec{v}_x \vec{v}_y \right) \right]^2
\;=\; \frac{1}{16} \left( |\rho_x \vec{v}_y + \rho_y \vec{v}_x|^2 - |\vec{v}_x \times \vec{v}_y|^2  \right)\,. \]
This is a polynomial equation for the matrix~$F_x F_y$, and thus the eigenvalues of the matrix
are the roots of the polynomial. This proves~(\ref{lpm}).

The identity~(\ref{ssign}) could be obtained from~(\ref{lpm}) by a direct calculation
using the Schwarz inequality and~(\ref{rhovc2}).
More directly, we can deduce from~(\ref{rhovc2}) that
\[ \det F_x \;=\; \rho_x^2 - |\vec{v}_x|^2 \;\leq\; 0 \:, \]
and thus
\[ \lambda_+ \lambda_- \;=\; \det \left( F_x F_y \right) \;=\;
 \det (F_x)\: \det (F_y) \;\geq\; 0\:. \]
 
\vspace*{-0.6cm}
\QED
This result raises the question whether a fermion system can be reconstructed
from the local correlation matrices. We first show that, for given~$\rho_x$ and~$\vec{v}_x$,
one can indeed construct a corresponding fermion system.
\begin{Prp} \label{prp42}
Suppose that for given parameters~$\rho_x \in \R$ and~$\vec{v}_x \in \R^3$
the relations~(\ref{rhovcond}, \ref{rhovc2}) are satisfied. Then there is a fermionic projector
which realizes~$\rho_x$ and~$\vec{v}_x$ as the corresponding local traces and
Bloch vectors, respectively.
\end{Prp}
{\Proof} For any~$x \in M$, (\ref{Fxd}) defines a local correlation matrix~$F_x$.
According to~(\ref{rhovc2}), $F_x$ has at most one positive and at most one negative
eigenvalue. Hence diagonalizing~$F_x$ by a unitary transformation~$U_x \in U(2)$
gives the representation~$F_x = U_x^{-1} D_x U_x$ with~$D_x={\mbox{diag}}(-\alpha,
\beta)$ and $\alpha,\beta \geq 0$. We define the fermion matrix by
\[ E_x \Psi \;=\; |D_x|^{\frac{1}{2}}\, U_x\:. \]
Then~(\ref{PsiFx}) is satisfied. Furthermore, the relations~(\ref{Fxd}) ensure that the
two columns of~$\Psi$ are orthonormal. Hence defining the fermionic projector by~(\ref{PsiP})
gives the desired fermion system.
\QED
However, the reconstruction of a fermion system from the local correlation matrices
is not unique, not even up to gauge transformations, as we now illustrate
by a simple example.
\begin{Example} \label{ex43} {\em{
In the case~$f=2$ and~$m=3$, we consider for a real parameter~$\alpha$
the family of fermion matrices
\[ E_1 \Psi \;=\; \left( \begin{matrix} 0 & 0 \\ 1 & 0 \end{matrix} \right)\:, \qquad
E_2 \Psi \;=\; \left( \begin{matrix} 0 & 0 \\ 0 & 1 \end{matrix} \right) \:,\qquad
E_3 \Psi \;=\; \left( \begin{matrix} \alpha & 1 \\ \alpha & 1 \end{matrix} \right) . \]
As required, the columns of~$\Psi$ are orthonormal vectors in~$H$. The corresponding
local traces and Bloch vectors are computed to be
\[ \rho_1 \;=\; \rho_2 \;=\; 1\:,\qquad \vec{v}_1 = - \vec{v}_2 = (0,0,1)^T\:,\qquad
\rho_3=0, \; \vec{v}_3=\vec{0}\:. \]
These quantities are independent of~$\alpha$. But the fermion systems 
corresponding to different values of~$\alpha$ clearly are not gauge equivalent.
\QEDrem
}} \end{Example}
In view of this example, it does not seem appropriate to restrict attention to
the local correlation matrices.
Instead, we shall describe the fermion system in discrete space-time by
the fermion matrix~(\ref{Psidef}), which is determined modulo the
transformations~(\ref{Psif}). The importance of the local correlation
matrices lies in the fact that the Bloch vectors will be very helpful for
illustrating the effect of spontaneous symmetry breaking.

\begin{figure}[t]
\begin{center}
\includegraphics[width=12cm]{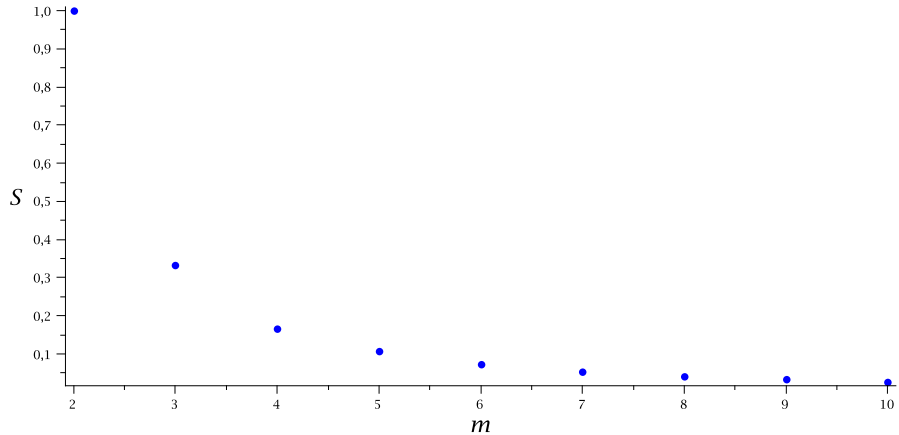}
\caption{Values of the minimal action for different numbers of space-time points}
\label{figMA}
\end{center}
\end{figure}

The variational principle in the critical case can be analyzed numerically
as a nonlinear optimization problem (for details see Appendix~\ref{appA}).
The minimum of the action as a function of the number of space-time points is shown
in Figure~\ref{figMA}.
In the following sections we discuss the different systems in detail, with the
emphasis on the symmetry structure of the minimizers.

\subsection{Two Space-Time Points} \label{sec42}
For two particles and two space-time points, the auxiliary variational principle
in the critical case is solved explicitly in~\cite[Example~3.2]{F1}.
To avoid repetitions, we here use a different method and work with the local
correlation matrices~(\ref{Fxdef}). According to~(\ref{rhovcond}),
the two matrices~$F_1$ and~$F_2$ have the form~(\ref{Fxd}) with
\beq \label{rv2}
\rho_1 + \rho_2 \;=\; 2 \:,\qquad \vec{v}_1 \;=\; -\vec{v}_2\:.
\eeq
Hence the term~$\vec{v}_x \times \vec{v}_y$ in Proposition~\ref{prp41}
vanishes,
\[ \lambda_\pm \;=\; \frac{1}{4} \Big[ \rho_x \rho_y + \vec{v}_x \vec{v}_y
\:\pm\: |\rho_x \vec{v}_y + \rho_y \vec{v}_x|  \Big] . \]
In particular, the~$\lambda_\pm$ are real. Furthermore, 
the identity~(\ref{ssign}) yields that~$\lambda_+$ and~$\lambda_-$ have the same sign.
Therefore, the Lagrangian~(\ref{Lsimp}) simplifies to
 \[ \Ll[A_{xy}] \;=\; \frac{1}{2}\, (\lambda_+ - \lambda_-)^2 \;=\;
\frac{1}{8}\;  \Bigl| \rho_x \vec{v}_y + \rho_y \vec{v}_x   \Bigr|  ^2\:.   \]
Summing over the space-time points and using that~$|\vec{v}_1|^2 = 
|\vec{v}_2|^2 = - \vec{v}_1 \vec{v}_2 =:v^2$, we obtain
\begin{eqnarray*}
\Ss &=& \frac{1}{8}\: \sum_{x=1}^2 \rho_x^2\: 2v^2 \:+\:
\frac{1}{4}\: \sum_{x,y=1}^2 \rho_x \rho_y\: v^2 \left(2 \delta_{xy} - 1 \right)
\:+\: \frac{1}{8}\: 2v^2 \sum_{y=1}^2 \rho_y^2 \\
&=& v^2 \left( \sum_{x=1}^2 \rho_x^2 \:-\: \frac{1}{4}\: \sum_{x,y=1}^2 \rho_x \rho_y \right)
\;=\; v^2 \left[ \rho_1^2+\rho_2^2 \:-\: 1 \right] ,
\end{eqnarray*}
where in the last step we used the first equation in~(\ref{rv2}).
Minimizing the action, we clearly want to choose~$v$ as small as possible.
Taking into account the constraint~(\ref{rhovc2}), we obtain
\[ \inf_v \Ss \;=\; \max(\rho_1^2, \rho_2^2) \left[ \rho_1^2+\rho_2^2 \:-\: 1 \right]\:. \]
By symmetry, we can assume that~$\rho_1 \geq \rho_2$. Using the first equation in~(\ref{rv2}),
we get
\[ \inf \Ss \;=\; \inf_{\rho_1 \in [1,\infty]} \rho_1^2 \left( 2 \rho_1^2 - 4 \rho_1 + 3 \right) . \]
The minimum is attained at~$\rho_1=1$. We conclude that for the minimizer the local
traces and the Bloch vectors have the form
\beq \label{rv22}
\rho_1 \;=\; \rho_2 \;=\; 1 \qquad {\mbox{and}} \qquad
|\vec{v}_1| \;=\; 1,\quad \vec{v}_2 \;=\; -\vec{v}_1 \:.
\eeq
The~$\lambda_\pm$ have the values
\begin{eqnarray}
{\mbox{for $A_{11}$ and~$A_{22}$}}: &\qquad&\lambda_+=1\:,\quad \lambda_-=0 \\
{\mbox{for $A_{12}$ and~$A_{21}$}}: &&\lambda_+= 0\:,\quad \lambda_-=0\:. \label{414}
\end{eqnarray}
Hence all points have timelike separation. The degeneracy in~(\ref{414}) means that we
are just in the boundary case between timelike and spacelike separation.

According to Proposition~\ref{prp42}, there exists a fermion system which
realizes~(\ref{rv22}). It is also easy to verify that the fermion matrix~$\Psi$
and the corresponding fermionic projector~$P$ as given by
\beq \label{PsiP2}
\Psi \;=\; \left( \begin{matrix}
0 & 0 \\ 1 & 0 \\ 0 & 0 \\ 0 & 1 \end{matrix} \right) , \qquad
P \;=\; \left( \begin{matrix}
0 & 0 & 0 & 0 \\ 0 & 1 & 0 & 0 \\ 0 & 0 & 0 & 0 \\ 0 & 0 & 0 & 1
\end{matrix} \right)
\eeq
indeed satisfy~(\ref{rv22}).  However, in view of Example~\ref{ex43}, the fermion system is in
general not determined by the local correlation matrices. 
Therefore, we now need to rely on~\cite[Example~3.2]{F1},
where it is shown that the fermionic projector corresponding to~(\ref{rv22})
is unique up to gauge transformations~(\ref{Pgt}) with~$U \in \G=U(1,1) \times U(1,1)$. Likewise, the fermion matrix in~(\ref{PsiP2})
is unique up to the transformations of the form~(\ref{Psif}).

The fermion matrix in~(\ref{PsiP2}) shows that the particle corresponding to the first column
is localized at the first space-time point, whereas the particle of the second column is localized at the
second space-time point. This configuration is what one would have expected in view of
the Pauli exclusion principle.
To verify that the fermion system~(\ref{PsiP2}) is permutation symmetric, we
choose the outer symmetry group~$\Oo=S_2=\{\1, \sigma\}$, where~$\sigma$ is the
transposition of the two space-time points. Setting 
\[ U(\1) \;=\; \1_{\C^4} \:,\qquad U(\sigma) \;=\; \left( \begin{matrix}
0 & \1_{\C^2} \\  \1_{\C^2} & 0 \end{matrix} \right) , \]
the relations~(\ref{USdef}) are satisfied. We conclude that the minimizing
fermion system for two space-time points has maximal outer symmetry, and no spontaneous
symmetry breaking occurs.

\subsection{Three Space-Time Points} \label{sec43}
For three space-time points, the critical action becomes so complicated 
that we could not compute the minimizers analytically. Numerically, we found that the
minimizing fermion systems are permutation symmetric. Assuming this permutation
symmetry, the fermion systems can also be treated analytically in closed form.
We first discuss the numerical results and then give the rigorous analysis of the
fermion systems with permutation symmetry.

\begin{figure}[t]
\begin{center}
 \includegraphics[width=5cm]{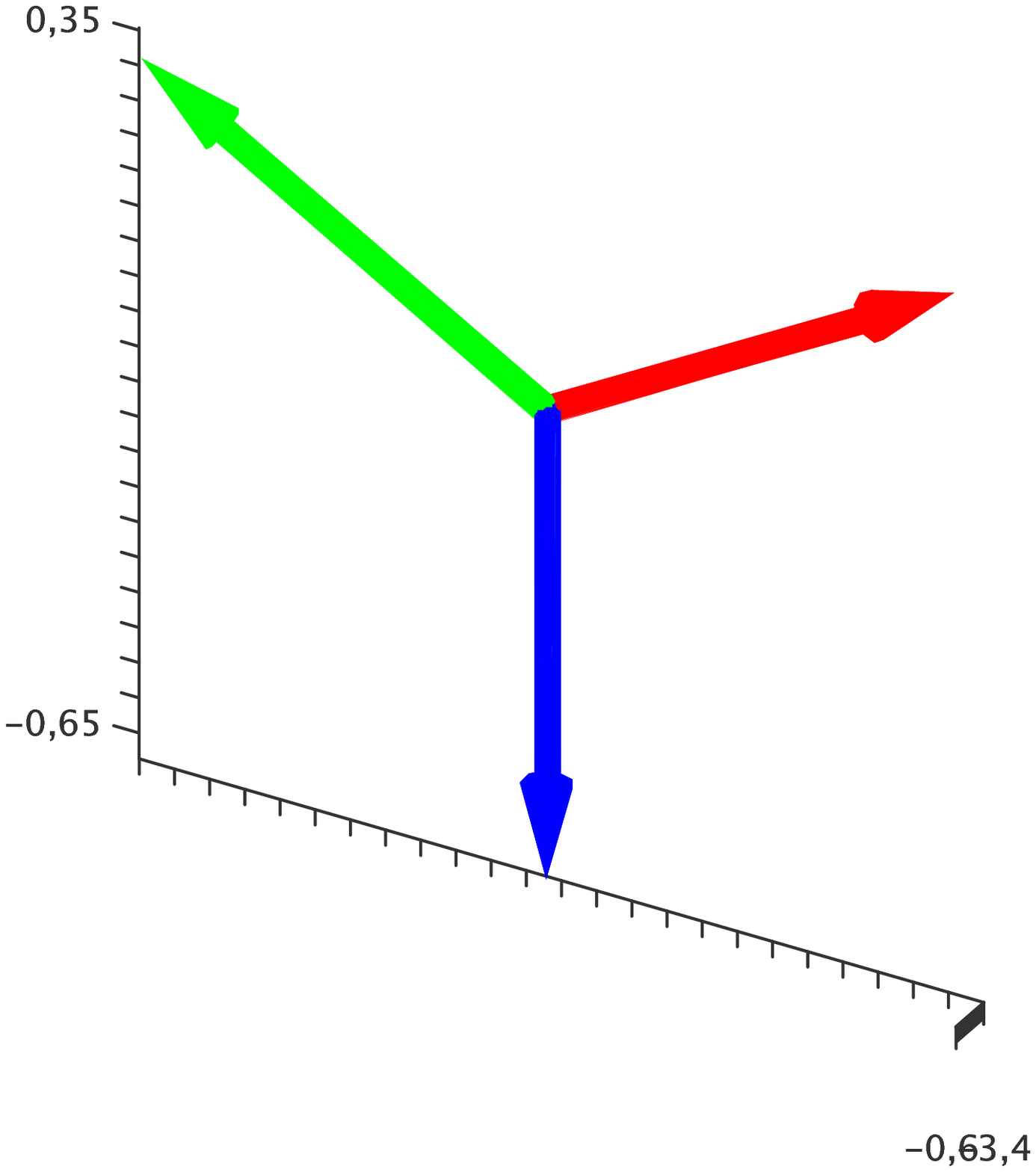}
 \includegraphics[width=5cm]{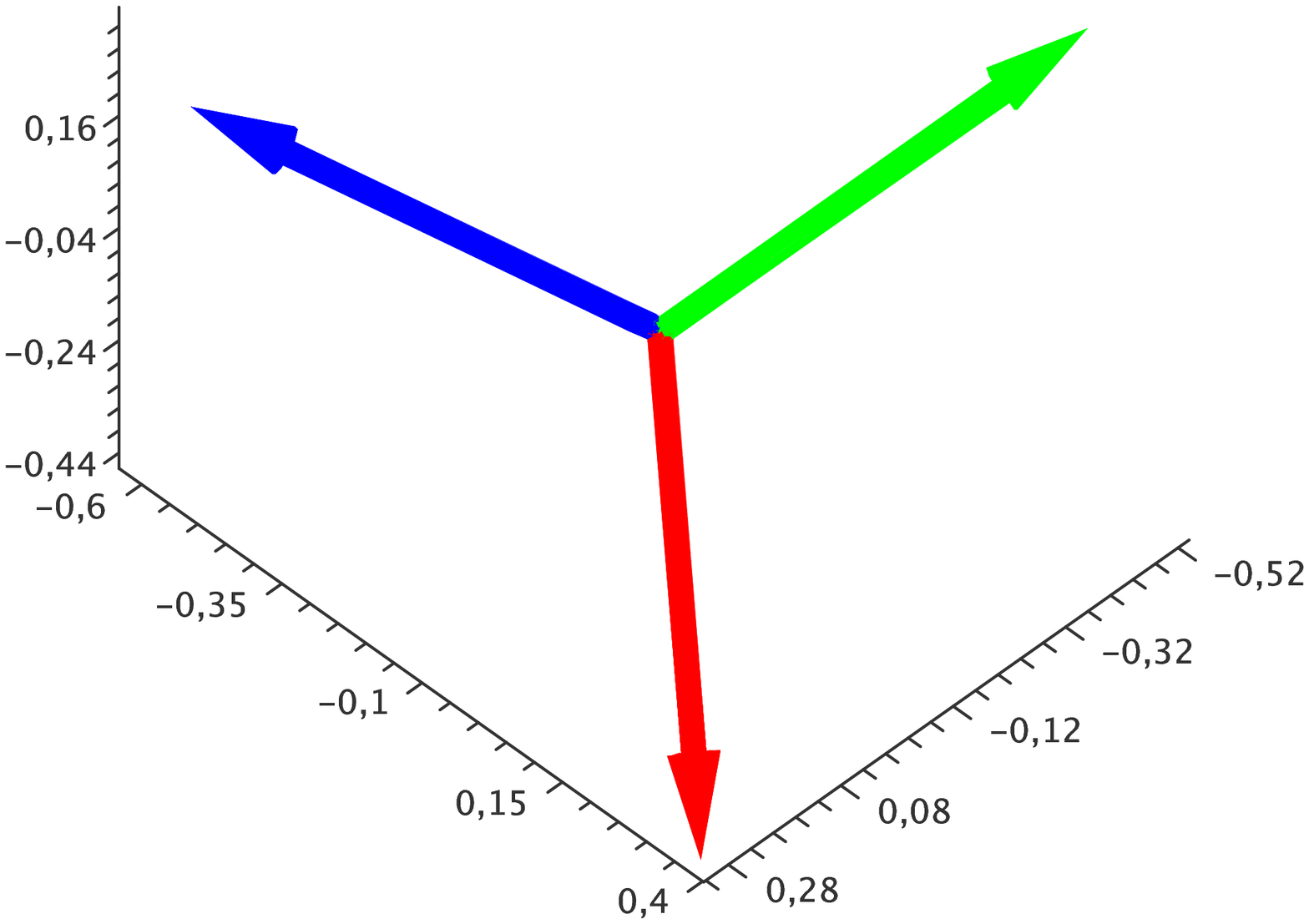}
\caption{Bloch vectors for three space-time points}
\label{figP3}
\end{center}
\end{figure}

The minimizers of our variational principle can be computed numerically
using methods of nonlinear optimization (for details see Appendix~\ref{appA}).
Computing the local correlation matrices~(\ref{Fxdef}, \ref{Fxd}) of the minimizers,
the resulting Bloch vectors~$\vec{v}_x$ form a plane equilateral triangle, see Figure~\ref{figP3}.
The local traces are all the same and coincide with the lengths of the Bloch vectors,
\[ |\vec{v}_x| \;=\; \rho_x \;=\; \frac{2}{3} \qquad \forall\, x \in \{1,2,3\}. \]
Since any two such equilateral triangles can be mapped onto each other by a rotation
in~$\R^3$, such that the numbering of the vertices is respected,
the Bloch vectors of the minimizers are unique up to the~$U(2)$-freedom~(\ref{U2},
\ref{Rot}). This suggests that the minimizing fermionic projector should be unique up
to gauge transformations.
Furthermore, any permutation of the labels of the vertices of the triangle can be
realized by a suitable rotation of~$\R^3$, and this indicates that the minimizer should
be permutation symmetric, i.e.\ that it should have the outer symmetry group~$\Oo=S_3$.
Indeed, our numerical solution is unique up to gauge transformations and is permutation
symmetric. We conclude that the minimizing fermion system for three
space-time points has maximal outer symmetry, and no spontaneous symmetry breaking
occurs.  Furthermore, all pairs of points are timelike separated, and thus no
non-trivial discrete causal structure appears.

In order to find closed expressions for the minimizer, we next characterize all fermion
systems with permutation symmetry using methods of~\cite{F3}.
\begin{Lemma} \label{lemma44}
In the case~$f=2$ and~$m=3$, for any~$\vartheta \geq 0$
the fermion system corresponding to the fermion matrix
\beq \label{fm}
\Psi \;=\; \frac{1}{\sqrt{6}}
\left( \begin{matrix} -2  \sinh \vartheta & 0 \\
0 & -2 \cosh \vartheta \\[.5em]
\sinh \vartheta & -\sqrt{3}  \sinh \vartheta \\
\sqrt{3} \cosh \vartheta & \cosh \vartheta \\[.5em]
\sinh \vartheta & \sqrt{3}  \sinh \vartheta \\
-\sqrt{3} \cosh \vartheta  & \cosh \vartheta
\end{matrix} \right)
\eeq
has the outer symmetry group~${\mathcal{O}}=S_3$.
Every fermionic projector with this outer symmetry group is gauge equivalent to
the fermionic projector corresponding to~(\ref{fm}) for some~$\vartheta \geq 0$.
The corresponding local density matrices~(\ref{Fxdef}) are of the form~(\ref{Fxd}) with
\begin{eqnarray}
\rho_1 &=& \rho_2 \;=\; \rho_3 \;=\; \frac{2}{3} \label{rho3} \\
\vec{v}_1 \;=\; v \left( \begin{matrix} 0 \\ 0 \\ -1 \end{matrix} \right) ,\qquad
\vec{v}_2 &=& v \left( \begin{matrix} \sqrt{3}/2 \\ 0 \\ 1/2 \end{matrix}
\right) ,\qquad
\vec{v}_3 \;=\; v \left( \begin{matrix} -\sqrt{3}/2 \\ 0 \\ 1/2 \end{matrix}
\right), \label{v3}
\end{eqnarray}
where
\beq \label{vdef}
v \;:=\; \frac{2}{3} \left(1 + \sinh^2 \vartheta \right) .
\eeq
\end{Lemma}
{\Proof} We point out that Definition~\ref{defouter} does not imply that the
mapping~$\sigma \mapsto U(\sigma)$ is a representation of~$\Oo$, and in
general this mapping cannot even be arranged to be a group representation
(this will be illustrated in the proof of Proposition~\ref{prp4st}).
However, for our fermion system and the group~$S_3$ one can indeed
arrange a group representation:
In~\cite[Proposition~9.4]{F3} it is shown that there exists a group
representation~$\sigma \mapsto U(\sigma)$ of the outer symmetry group~$\Oo=S_3$
provided that for the set~${\mathcal{T}} \subset \Oo$ of all transpositions the following
conditions hold:
\begin{description}
\item[(A)] $U(\tau)^2=\1$ for all~$\tau \in {\mathcal{T}}$.
\item[(B)] For all~$\tau, \tau' \in {\mathcal{T}}$ we have the implication
\[ [\tau, \tau']=\1 \quad \Longrightarrow \quad
[U(\tau), U(\tau')]=\1\:. \]
\item[(C)] For all distinct~$x, y, z \in M$,
\[ U(\tau_{x,y})\, U(\tau_{y,z})\, U(\tau_{x,y}) \;=\; \pm U(\tau_{x,z})\:, \]
where~$\tau_{xy}$ denotes the transposition of~$x$ with~$y$.
\end{description}
For the group~$S_3$, there are no commuting transpositions, and thus~(B) is
trivially satisfied. To satisfy~(A) and~(C) we proceed as in~\cite[proof of Theorem~9.1]{F3}.
We can assume that our fermion system is a simple system (see~\cite[Section~4]{F3}).
According to~\cite[Lemma~9.5]{F3}, $U(\tau)$ is unique up to a phase.
By choosing the phase appropriately, we can arrange~(A); this fixes the operators~$U(\tau)$
up to a sign. Furthermore, we know that~(C) holds with a general phase, i.e.\
for all distinct~$x, y, z \in M$,
\[ U_{x,y}\, U_{y,z}\, U_{x,y}\, U_{x,z} \;=\;
e^{i \vartheta(x,y,z)}\:\1 \:, \]
where~$U_{x,y} \equiv U(\tau_{xy})$. The sign of the phase factor depends on our arbitrary choice
of the signs of the operators~$U(\tau)$. But up to the sign,
the factor~$e^{i \vartheta(x,y,z)}$ is uniquely defined. From the permutation
symmetry of our system
we conclude that it is a constant independent of the space-time points, i.e.
\beq \label{Ufixed}
U_{x,y}\, U_{y,z}\, U_{x,y}\, U_{x,z} \;=\;
\pm e^{i \vartheta}\: \1 \qquad
{\mbox{for all distinct~$x, y, z \in M$}}\:.
\eeq
Multiplying from the right by~$U_{x,z}$ and from the
left by~$U_{x,y} U_{y,z} U_{x,y}$, we get
the same relation, but with the sign of~$\vartheta$ reversed.
We conclude that~$e^{i \vartheta}=\pm 1$, proving~(C). Hence we can arrange that the mapping $\sigma \mapsto U(\sigma)$ is a group representation.

According to~\cite[Lemma~2.10]{F3}, $H$ can be decomposed into
an orthogonal direct sum of definite subspaces, which are irreducible
under the group representation~$\sigma \mapsto U(\sigma)$.
Taking the direct sum of all positive irreducible subspaces gives
a subspace~$I^+ \subset H$. Introducing~$I^-$ similarly, we
obtain the decomposition~$H = I^+ \oplus I^-$. Hence these
subspaces are both three-dimensional, and we can arrange by
a gauge transformation that 
\[ E_x I^+ \;=\; \left( \begin{matrix} \C \\ 0 \end{matrix} \right)
\quad {\mbox{and}} \quad
E_x I^- \;=\; \left( \begin{matrix} 0 \\ \C \end{matrix} \right)
\qquad \forall\, x \in M . \]

In~$I^-$ we choose a basis~$(e_x)_{x \in M}$ such that~$E_x e_y = \delta_{xy}
\, (0,1)^T$ and write a vector $v \in I^-$ in components as~$v=(v_x)_{x \in M}$.
Then the right equation in~(\ref{USdef}) yields that~$U$
is a permutation up to a phase, i.e. for every~$v \in I^-$,
\[ U(\sigma)\, v_x \;=\; e^{-i \alpha(x,\sigma)}\: v_{\sigma(x)}\:. \]
Using the gauge freedom together with the fact that~$U$ is a group representation,
we can arrange that~$U$ is of one of the following two forms\footnote{
An easy method to see that these two representations are not gauge equivalent
is to compute the determinant of~$U(\sigma)$ for an odd permutation~$\sigma$.
In cases~(i) and~(ii) this gives plus and minus one, respectively.
Clearly, the determinant of~$U(\sigma)$ does not change under gauge transformations
$U(\sigma) \rightarrow V U(\sigma) V^{-1}$ with~$V \in U(3)$.}:
\begin{description}
\item[(i)] $U(\sigma)\, v_x \;=\; v_{\sigma(x)}$
\item[(ii)] $U(\sigma)\, v_x \;=\; \sgn(\sigma) \: v_{\sigma(x)}$
\end{description}
In both cases, the irreducible subspaces  of these representations
(considered as representations on~$\C^3$, which we can identify
via our basis with~$I^-$) are
\[ J_1^- \;=\; \C\, \left( \begin{matrix} 1 \\ 1 \\ 1 \end{matrix} \right) 
\qquad {\mbox{and}} \qquad J_2^- \;=\; (J_1^-)^\perp\: . \]
On $I^+$ we can proceed exactly as for~$I^-$. Thus our representation~$U$
restricted to $I^+$ is again of the form~(i) or~(ii). We denote the irreducible subspaces
by~$J_1^+, J_2^+ \subset I^+$.

Clearly, $U$ acts trivially on~$J_1:=J_1^+ \oplus J_1^-$, which is
an inner product space of signature~$(1,1)$. Since the image of~$P$
must be negative definite, $J_1$ can be occupied by at most one particle.
On~$J_2:=J_2^+ \oplus J_2^-$, $U$ acts as two copies of the same $2$-dimensional
irreducible representation. According to Schur's lemma, the range of~$P|_{J_2}$
must be even. Since~$J_2$ has signature~$(2,2)$, it can be occupied by zero
or two particles. Thus the only way to build up a fermionic projector of rank two
is to occupy~$J_1$ by zero and~$J_2$ by two particles.

The representations of~$U$ on~$J_2^+$ and~$J_2^-$
are clearly equivalent. But the subtle point is that the representation matrices are
the same only if in~$J_2^+$ and~$J_2^-$ we choose appropriate bases.
If~$U$ restricted to~$I^+$ and~$I^-$ is of the same form~(i) or~(ii), we can
choose the same orthonormal basis in~$J_2^+$ and~$J_2^-$, for example the two vectors
\[ u \;=\; \frac{1}{\sqrt{2}} \left( \begin{matrix} 0 , 1 , -1 \end{matrix} \right)^T
\:, \qquad v \;=\; \frac{1}{\sqrt{6}} \left( \begin{matrix} -2 , 1 , 1 \end{matrix} \right)^T . \]
This basis gives rise to the factorization
\beq \label{fact1}
J \;=\; \C^2 \otimes V \:,
\eeq
where the components of~$\C^2$ are the coefficients of the basis vectors~$u$ and~$v$,
and~$V$ is an inner product space of signature~$(1,1)$.
The representation~$U$ also factors,
\beq \label{fact2}
U|_{J} \;=\; U_{\mbox{\scriptsize{irr}}} \otimes \1 \:,\spc
J \;=\; \C^2 \otimes V \:,
\eeq
where~$U_{\mbox{\scriptsize{irr}}}$ is an irreducible matrix representation on~$\C^2$.
According to Schur's lemma, the fermionic projector must be trivial on the first factor,
\[ P|_{J} \;=\; \1_{\C^2} \otimes p\:, \]
where~$p$ is a projector onto a one-dimensional negative definite subspace of~$V$.
Representing the unit vector in the image of~$p$ as~$(\sinh \vartheta, \cosh \vartheta)$,
we obtain the fermion matrix
\[ E_x \Psi \;=\; \left( \begin{matrix} u_x \,\sinh \vartheta  & v_x \,\sinh \vartheta \\
u_x \,\cosh \vartheta  & v_x \,\cosh \vartheta \end{matrix} \right) \:. \]
This fermion matrix is gauge equivalent to~(\ref{fm}) with~$\vartheta=0$.
In the case that~$U$ is of a different form on~$I^+$ and~$I^-$ (i.e.\ that $U$
restricted to~$I^+$  is of the form~(i) and~$U$ restricted to~$I^-$ is of the
form~(ii), or vice versa), we can arrange that the representation matrices
on~$I^+$ and~$I^-$ are the same by choosing in~$I^+$ the basis $(u,v)$
and in~$I^-$ the basis $(v,-u)$. Then we again have the factorizations~(\ref{fact1},
\ref{fact2}), and we thus obtain the fermion matrix
\[ E_x \Psi \;=\; \left( \begin{matrix} v_x \,\sinh \vartheta  & -u_x \,\sinh \vartheta \\
u_x \,\cosh \vartheta  & v_x \,\cosh \vartheta \end{matrix} \right) \:. \]
This coincides precisely with~(\ref{fm}).

The local traces and Bloch vectors follow from a straightforward calculation.
\QED
For permutation symmetric systems, we can easily analyze the critical case of
the variational principle analytically.
\begin{Prp}\label{prp45} Considering the critical variational principle in the class of
fermionic projectors with outer symmetry group~$\Oo=S_3$, the minimum is
attained by the fermionic projector corresponding to the fermion matrix~(\ref{fm})
with~$\vartheta=0$. The minimizing fermionic projector is unique up to
gauge transformations.
\end{Prp}
{\Proof} We compute the~$\lambda_\pm$ using Proposition~\ref{prp41}
together with Lemma~\ref{lemma44}. For the roots of the characteristic
polynomial of~$A_{xx}$ we obtain
\beq \label{lampm1}
\lambda_\pm \;=\; \frac{1}{4} \left[ \frac{4}{9} + v^2 \pm \frac{4}{3}\: v \right] .
\eeq
For any~$x \neq y$, the Bloch vectors in~(\ref{v3}) have an angle of~$120^\circ$, and thus
\beq \label{lampm2}
\lambda_\pm \;=\; \frac{1}{4} \left[ \frac{4}{9} - \frac{v^2}{2} \pm v\:
\sqrt{\frac{4}{9} - \frac{3}{4}\: v^2 } \right] .
\eeq
A straightforward computation yields that
\beq \label{actionm3} \Ss \;=\; \frac{2}{3}\, v^2 + \Theta\!\left( \frac{16}{27} - v^2 \right)
\left( \frac{v^2}{3} - \frac{9}{16}\: v^4 \right) . \eeq
According to~(\ref{vdef}), we minimize the action under the constraint~$v \geq 2/3$.
A short calculation shows that the minimum is attained at~$v=2/3$, giving the result.
\QED

\subsection{Four Space-Time Points: Spontaneous Breaking of the Parity Symmetry}
\begin{figure}[t]
\begin{center}
 \includegraphics[width=5cm]{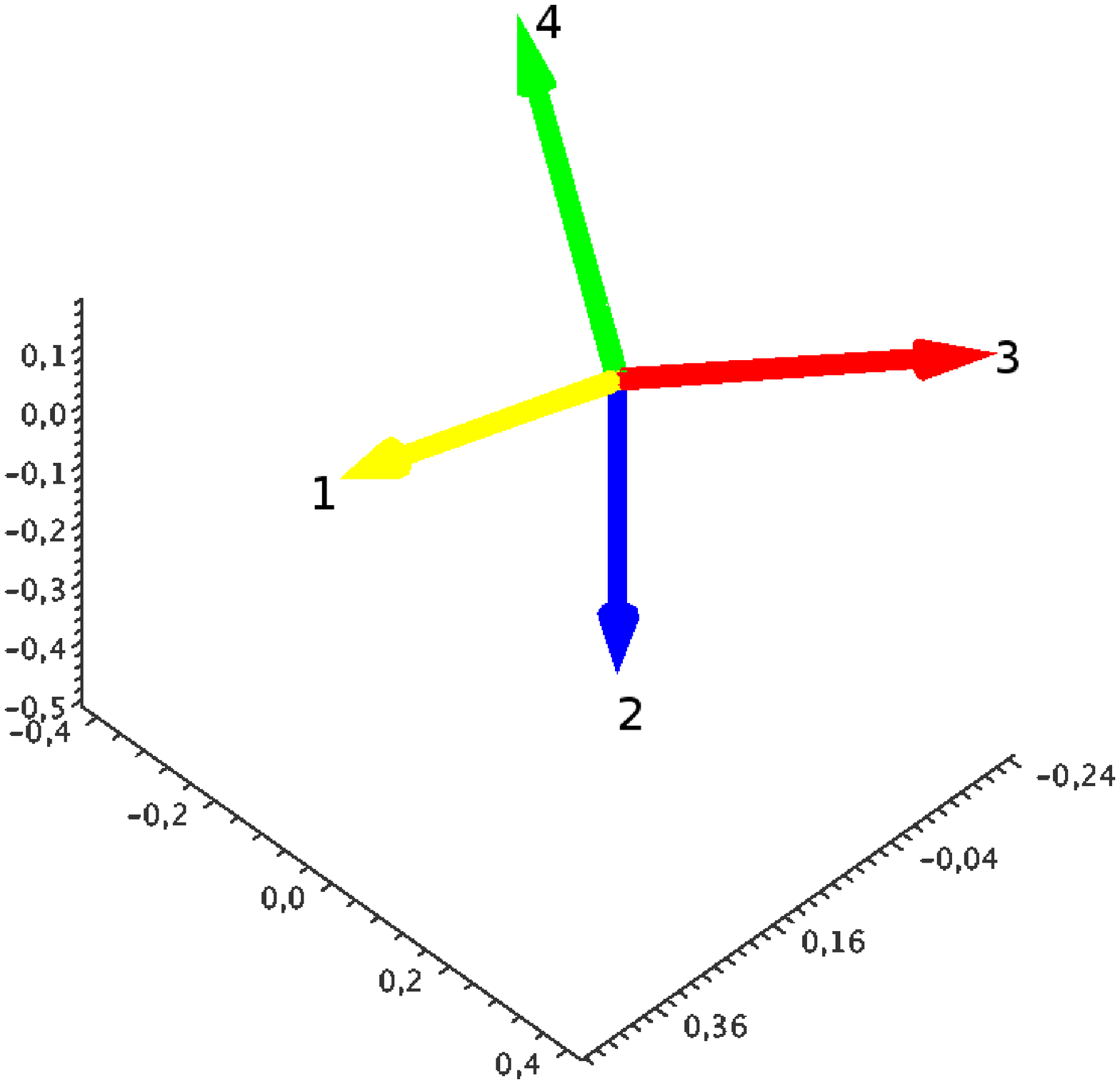}
 \includegraphics[width=5cm]{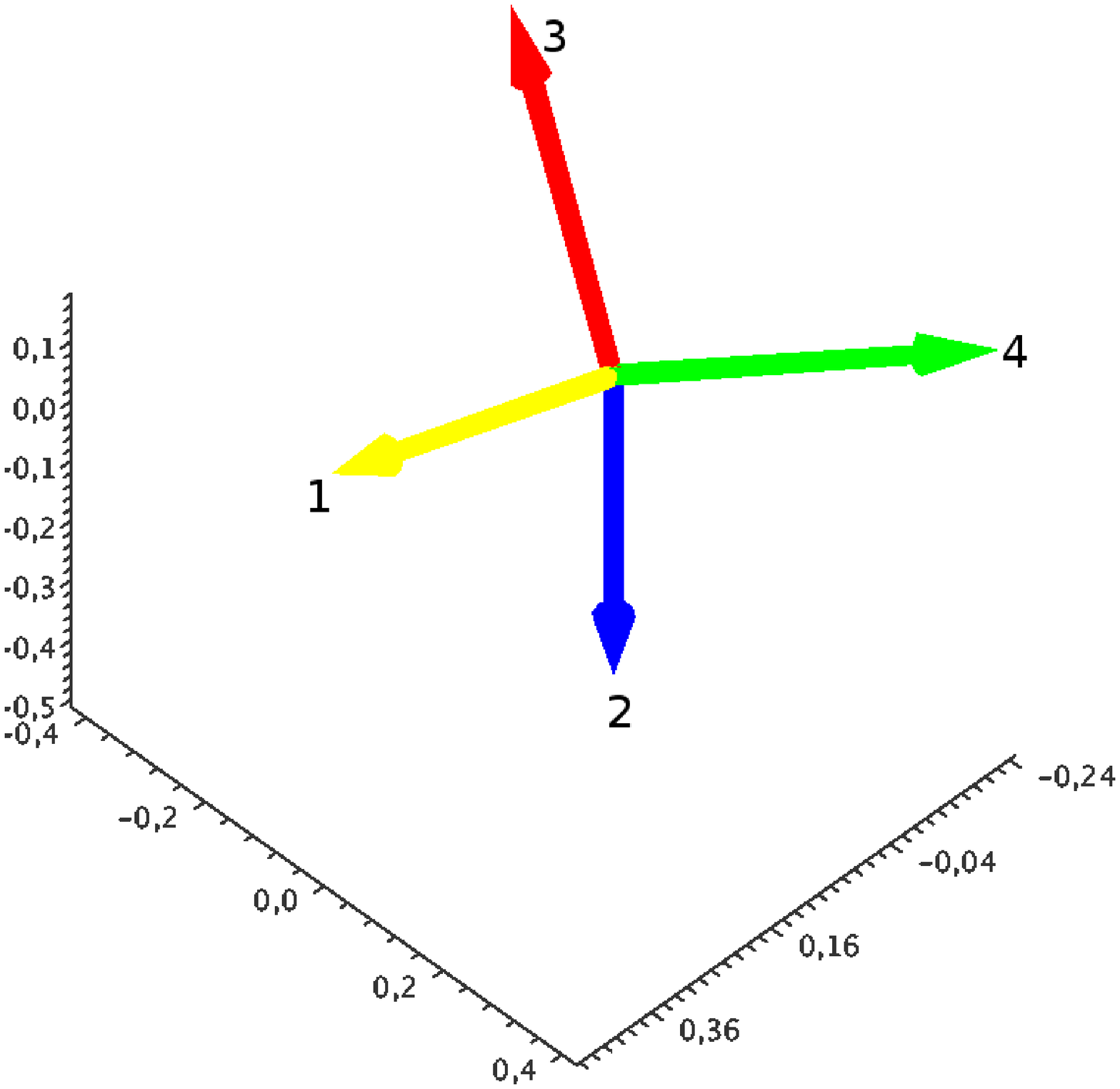}
 \caption{Bloch vectors for four space-time points} \label{figtetra}
\end{center}
\end{figure}

In the case~$f=2$ and~$m=4$, Theorem~\ref{sym} yields that there are no
fermion systems with outer symmetry group~$\Oo=S_4$. In other words, the
permutation symmetry is spontaneously broken. This effect can be seen in our
numerics as follows. Similar as for three space-time points, the lengths of the
Bloch vectors are all equal and coincide with the local traces,
\beq \label{loctr4}
|\vec{v}_x| \;=\; \rho_x \;=\; \frac{1}{2} \qquad \forall\, x \in \{1,\ldots,4\}.
\eeq
The Bloch vectors of our numerical minimizers
form a tetrahedron, see Figure~\ref{figtetra}.
By labeling its vertices by the corresponding space-time points $1,\ldots,4$,
we give the tetrahedron an orientation. As a consequence, two such labeled tetrahedra can
in general not be mapped onto each other by a rotation in~$\R^3$, as
one sees in the two examples of Figure~\ref{figtetra}.
In other words, there are two minimizing fermionic projectors, which distinguish a left or right orientation on
the tetrahedrons and are not gauge equivalent.
Since the parity transformation~$\vec{x} \rightarrow -\vec{x}$ reverses the orientation
of a tetrahedron, we can also say that the symmetry
under parity transformations is spontaneously broken by the minimizing fermionic projector.
The outer symmetry group of a minimizer is merely the alternating group~$A_4$ of all
even permutations, because only those permutations can be realized by rotations
in~$\R^3$. This fact can also be expressed by saying that the minimizing fermionic projector
spontaneously breaks the permutation symmetry of the space-time points, giving
rise to an orientation of the space-time points.

In the next proposition we give closed formulas for $A_4$-symmetric  
fermion systems
satisfying~(\ref{loctr4}). These fermion systems coincide precisely  
with our
numerically found minimizers.
Using the methods of~\cite{F3}, one could characterize all fermionic  
projectors
with outer symmetry group~$A_4$. However, since the group~$A_4$ and its
central extensions have many irreducible representations, this  
analysis would be
lengthy. For simplicity, we here only give the fermion matrices, but  
we do not prove uniqueness.
\begin{Prp} \label{prp4st}
For every~$\varphi =\pm 2 \pi/3$, the fermion system corresponding to  
the fermion matrix
\beq \label{fm4st}
E_x \Psi \;=\; \psi_x \otimes \left( \begin{matrix} 0 \\ 1 \end
{matrix} \right)
\;\equiv\; \left( \begin{matrix} 0 \;\;\;\; 0 \\
\psi_x \end{matrix} \right) \quad {\mbox{with}} \quad
\left( \begin{matrix} \psi_1 \\ \psi_2 \\ \psi_3 \\ \psi_4 \end
{matrix} \right)
\;=\; \frac{1}{\sqrt{6}}
\left( \begin{matrix} \sqrt{3} & 0 \\ 1 & \sqrt{2} \\ 1 & \sqrt{2}\,  
e^{i \varphi} \\
1 & \sqrt{2}\, e^{-i \varphi} \end{matrix} \right)
\eeq
has the outer symmetry group~$A_4$. The local correlation matrices~(\ref
{Fxdef}, \ref{Fxd})
satisfy~(\ref{loctr4}). The corresponding Bloch vectors are in a  
tetrahedron configuration.

The Bloch vectors of the two systems corresponding to~$\varphi=2 \pi/3$
and~$\varphi=-2 \pi/3$ have opposite orientation. The corresponding  
fermionic projectors
are not gauge equivalent.
\end{Prp}
{\Proof} The group~$A_4$ is generated by the two even permutations
\[ \sigma \;=\; \left( \begin{matrix} 1 & 2 & 3 & 4 \\
1 & 3 & 4 & 2 \end{matrix}\right) \qquad {\mbox{and}} \qquad
\tau \;=\; \left( \begin{matrix} 1 & 2 & 3 & 4 \\
2 & 1 & 4 & 3 \end{matrix}\right) . \]
Hence it suffices to satisfy the relations~(\ref{USdef}) for these two group elements.
Since all transformations will leave the first component of~$E_x \Psi$
at each space-time point unchanged, we only consider the  
transformation of
the vector~$\psi = (\psi_1, \ldots, \psi_4)^T$, on which the inner  
product induced
by~$H$ is the standard scalar product on~$\C^4$. We introduce the two  
unitary
matrices
\beq \label{Ust}
U(\sigma) \;=\; \left( \begin{matrix} 1 & 0 & 0 & 0 \\
0 & 0 & 0 & 1 \\
0 & 1 & 0 & 0 \\
0 & 0 & 1 & 0 \end{matrix} \right) , \qquad
U(\tau) \;=\; \left( \begin{matrix} 0 & 1 & 0 & 0 \\
1 & 0 & 0 & 0 \\
0 & 0 & 0 & i \\
0 & 0 & -i & 0
\end{matrix} \right) .
\eeq
A direct computation using the trigonometric identities
\[  1+2 e^{\pm \frac{2 \pi i}{3}} \;=\;  \pm i \sqrt{3}\:, \qquad
1-e^{\pm \frac{2 \pi i}{3}} \;=\; \sqrt{3}\; e^{\mp \frac{2 \pi i}
{3}} \]
yields that
\beq \label{psit}
U(\sigma) \,\psi \;=\; \psi \,V(\sigma)  \:, \qquad
U(\tau) \,\psi \;=\; \psi \,V(\tau) \:,
\eeq
where~$V$ are the following $U(2)$-matrices:
\[ V(\sigma) \;=\; \left( \begin{matrix} 1 & 0 \\ 0 & e^{i \varphi}  
\end{matrix}
\right)  \:, \qquad
V(\tau) \;=\; \frac{1}{\sqrt{3}} \left( \begin{matrix} 1 & \sqrt{2} \\ \sqrt{2} & -1 \end{matrix}
\right) .  \]
Substituting~(\ref{psit}) into the formula for the fermionic  
projector~(\ref{PsiP}),
the matrices~$V$ drop out, and we obtain the left equation in~(\ref
{USdef}).
The right equation is obvious from~(\ref{Ust}). We conclude that~$A_4
$ really is
an outer symmetry group of the fermion system with fermion matrix~
(\ref{fm4st}).

It can be verified by a straightforward computation that the Bloch  
vectors corresponding
to~(\ref{fm4st}) form a tetrahedron and satisfy~(\ref{loctr4}).  
Alternatively, these
facts are verified without a computation as follows. Since~$A_4$ acts  
transitively
on~$M$, the local traces~$\rho_x$ all coincide. From~(\ref{rhovcond}) we
conclude that~$\rho_x=1/2$. Next, it is obvious from~(\ref{fm4st})  
that the
matrix~$E_x \Psi$ has rank one. Using~(\ref{PsiFx}) it follows that the
local correlation matrices~$F_x$ all have rank at most one. This  
means in the
representation~(\ref{Fxd}) that~$|\vec{v}_x|=\rho_x$.
Hence the Bloch vectors~$\vec{v}_x$ have non-zero length and are in an
$A_4$-symmetric configuration. The only such configuration is a  
tetrahedron.

As is obvious from~(\ref{fm4st}), the systems corresponding to~$
\varphi=2 \pi/3$
and~$\varphi=-2 \pi/3$ are obtained from each other by exchanging the  
third and fourth
space-time point. Since this transformation changes the orientation  
of the tetrahedron,
the tetrahedra of the two systems cannot be mapped onto each other by  
a rotation in~$\R^3$.
However, the local correlation matrices of gauge equivalent fermionic  
projectors
are unitarily equivalent, because~(\ref{Fxdef}) is gauge invariant  
and depends only
on the choice of the basis in~$P(H)$. We conclude that the fermionic  
projectors
corresponding to~$\varphi=\pm 2 \pi/3$ cannot be gauge equivalent.
\QED

The $A_4$-symmetric fermion systems of the previous proposition are also
interesting from the point of view of representation theory,
because they are examples where the outer symmetry group has no
representation. In order to get a group representation, one needs to construct a central
extension~$\hat{\mathcal{N}}$ of the outer symmetry group.
In the next proposition we construct this group extension for the systems
of Proposition~\ref{prp4st}. This explains why in Definition~\ref{defouter}
it was important {\em{not}} to demand that the mapping~$\sigma \mapsto U(\sigma)$
should be a representation of~$\Oo$. The following example also illustrates the
abstract constructions in~\cite[Section~5]{F3}.
We denote the cyclic group of order two by~$\Z_2$ (e.g.\ one can take
$\Z_2=\{1,-1\}$ with the standard multiplication).
By a $\Z_2$-extension of~$A_4$ we mean a group~$\hat{\mathcal{N}}$
together with group homomorphisms~$\iota$ and~$\pi$ such that the following sequence
is exact,
\beq \label{exact}
0 \;\longrightarrow\; \Z_2 \;\stackrel{\iota}{\longrightarrow}\; \hat{\mathcal{N}}
\;\stackrel{\pi}{\longrightarrow}\; A_4 \;\longrightarrow\; 0 \:,
\eeq
where $0$ denotes the trivial group.
\begin{Prp} There is no unitary representation~$U(\sigma)$ of the outer symmetry
group~$A_4$ which satisfies~(\ref{USdef}).
There exists a $\Z_2$-extension~$\hat{\mathcal{N}}$ of~$A_4$
together with a unitary representation~$\hat{U}$
which describes the outer symmetry in the sense that for all~$\sigma \in \hat{\mathcal{N}}$,
\beq \label{osext}
\hat{U}(\sigma)\,P\hat{U}\,(\sigma)^{-1} \;=\; P \quad\;\; {\mbox{and}} \quad\;\;
\hat{U}(\sigma)\, E_x \,\hat{U}(\sigma)^{-1} \;=\; E_{(\pi(\sigma))(x)} \quad
\forall\, x \in \{1,\ldots,4\}\:,
\eeq
where~$\pi\::\: \hat{\mathcal{N}} \rightarrow A_4$ is the projection in~(\ref{exact}).
\end{Prp}
{\Proof} A short computation shows that the conditions~(\ref{USdef}) determine the
matrices in~(\ref{Ust}) uniquely up to a phase. (Using the notion
from~\cite{F3}, this could be expressed by saying that the free gauge group is
merely the~$U(1)$ of global phase transformations.) In order to obtain a group representation,
we must certainly satisfy the conditions~$U(\sigma)^3=\1$ and~$U(\pi)^2=\1$. Hence~$U(\sigma)$
must coincide with the matrix in~(\ref{Ust}) up to a sign, whereas~$U(\tau)$
could be modified by multiplication with~$e^{\pm 2 \pi i/3}$. Computing the product
$U(\tau) U(\sigma)$, we find that
\[ (U(\tau) U(\sigma))^3 \;=\; \pm i \1\:. \]
If~$U$ were a representation, this would be equal to~$U((\tau \sigma)^3)=U(\1)=\1$,
a contradiction.

In order to construct $\hat{\mathcal{N}}$, we first note that changing the phase of~$U$
changes the phase of the matrix~$V$ in~(\ref{psit}) accordingly. But since~$V$ is
only a $(2 \times 2)$-matrix, we can fix its phase up to a sign
by imposing that its determinant should be equal to one. This leads us to introduce the matrices
\[ V^\pm(\sigma) \;=\; \pm
\left( \begin{matrix} e^{-\frac{i \varphi}{2}} & 0 \\ 0 & e^{\frac{i \varphi}{2}} \end{matrix}
\right)  \:, \qquad
V^\pm(\tau) \;=\; \pm
\frac{i}{\sqrt{3}} \left( \begin{matrix} 1 & \sqrt{2} \\ \sqrt{2} & -1 \end{matrix}
\right)  \]
and to define the matrices~$U^\pm(\tau)$ and~$U^\pm(\sigma)$ by the relations
\[ U^\pm(\sigma) \,\psi \;=\; \psi \,V^\pm(\sigma)  \:, \qquad
U^\pm(\tau) \,\psi \;=\; \psi \,V^\pm(\tau) \:. \]
The four matrices $\{U^\pm(\sigma),U^\pm(\tau)\}$ clearly satisfy the left equation in~(\ref{USdef}).
They generate a group, denoted by~$\hat{\mathcal{N}}$.
The injection~$\iota$ in~(\ref{exact}) is defined by~$\iota \::\: \Z_2 \rightarrow \hat{\mathcal{N}} \::\: \pm 1 \mapsto \pm \1$.
Every matrix~$U \in \hat{\mathcal{N}}$ clearly satisfies the
right relation in~(\ref{USdef}) for a suitable~$\sigma \in A_4$. This defines the
mapping~$\pi$ in~(\ref{exact}). Then the relations~(\ref{osext}) are satisfied by
construction. Obviously, $\iota$ is injective, and its image is the kernel of~$\pi$.
To verify that~$\pi$ is surjective, one represents any group element~$g \in A_4$
as a product of powers of $\sigma$ and~$\tau$. Taking the product of the powers
of the corresponding matrices~$U^+(\sigma)$ and~$U^+(\tau)$ gives
a matrix~$U \in \hat{\mathcal{N}}$ with~$\pi(U)=g$.
\QED

\subsection{Five Space-Time Points: Spontaneous Breaking of the Translation Symmetry}
\begin{figure}[t]
\begin{center}
 \includegraphics[width=5cm]{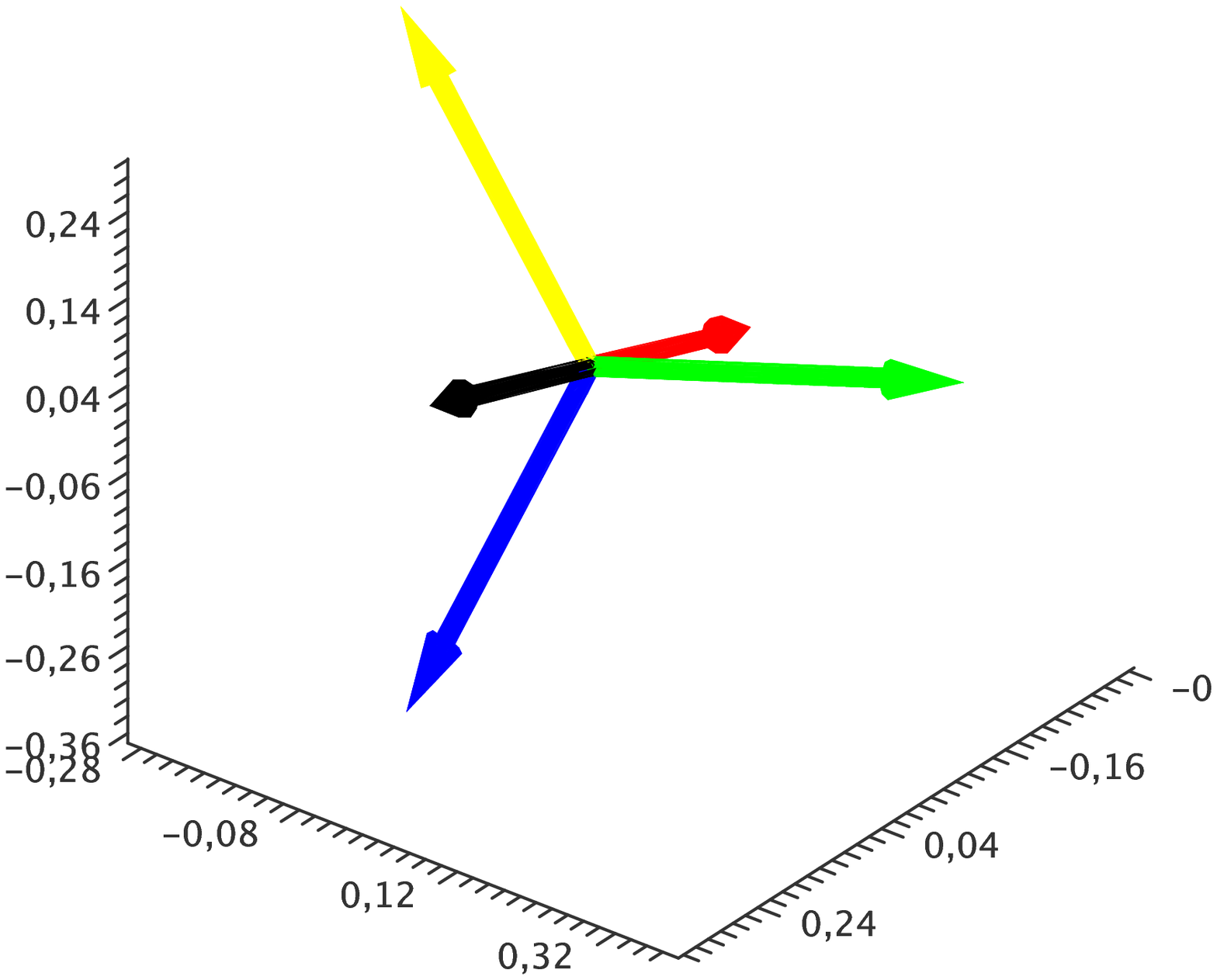}
 \includegraphics[width=5cm]{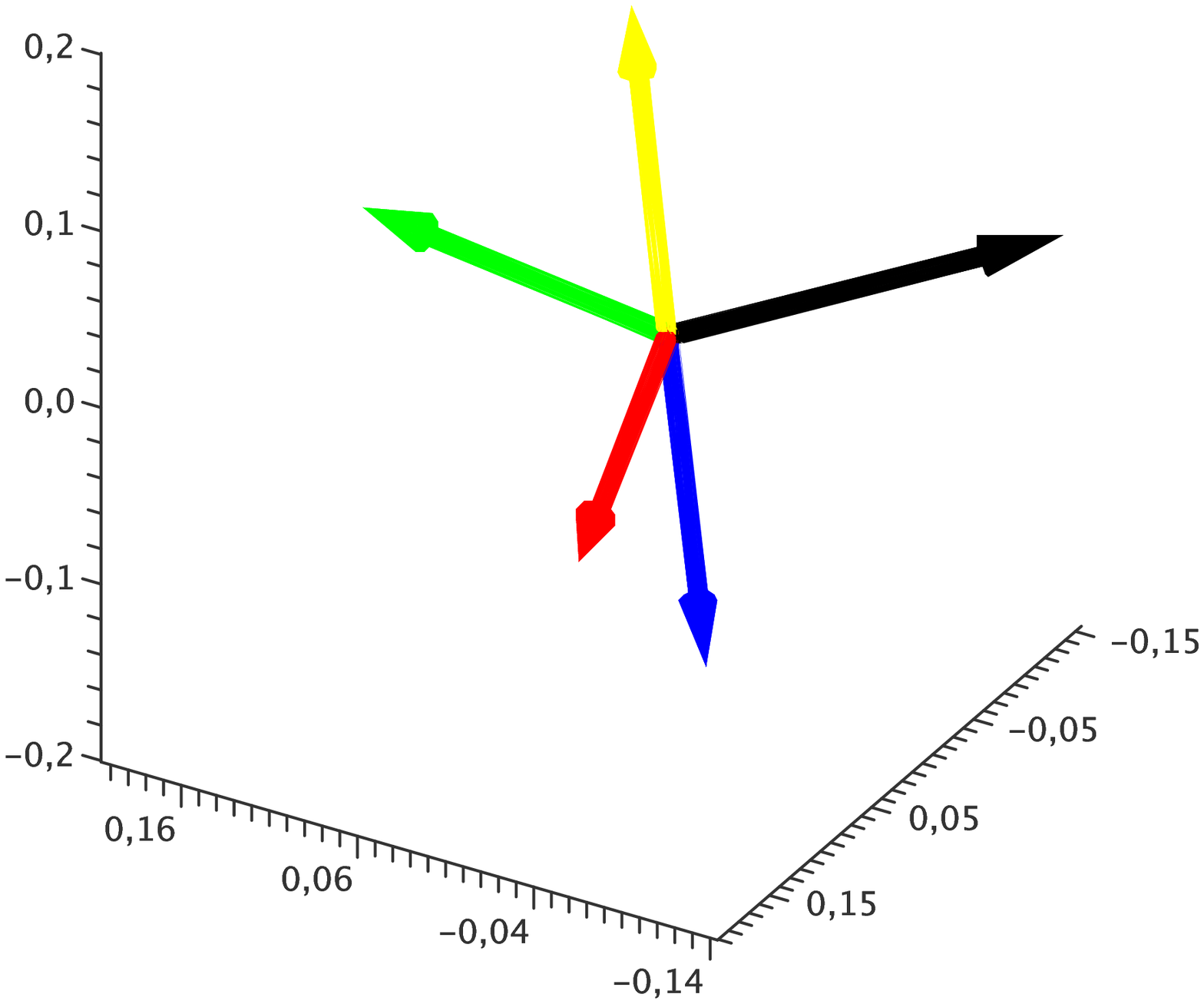}
 \caption{Bloch vectors for five space-time points}
 \label{figPV5}
\end{center}
\end{figure}

For five space-time points, we obtained the following numerical results.
The Bloch vectors of the minimizers are unique
up to rotations in space and a different numbering of the vertices; Figure~\ref{figPV5}
shows a typical example.
The lengths of the Bloch vectors coincide with the corresponding local traces,
\beq \label{vrc5}
|\vec{v}_x| \;=\; \rho_x \qquad \forall\, x \in \{1,\ldots, 5\}\:.
\eeq
A new feature, which we found only for five space-time points, is that the
local traces, and thus also the lengths of the Bloch vectors, have different values.
Namely, $\rho_x \approx 0.3883$ for two of the space-time points, whereas
$\rho_x \approx 0.4077$ for the other three space-time points.
The two shorter Bloch vectors lie on a straight line, whereas
the other three vectors lie in a plane perpendicular to this line and form an equilateral
triangle. The subgroup of rotations which leave the Bloch vectors unchanged
is the group~$S_3$, corresponding to the three rotations about the straight line
by the angles~$0$ and~$\pm 120^\circ$ and the three rotations by $180^\circ$ around the three Bloch vectors
in the perpendicular plane. This group is indeed the outer symmetry group.
Hence the symmetric group~$S_5$ is spontaneously broken to the much smaller
group~$\Oo=S_3$. Again, all space-time points have timelike separation, and so
the emerging discrete causal structure is trivial.

The spontaneous breaking of the outer symmetry group to the group~$S_3$ can be interpreted
as follows. As usual, we call a space-time homogeneous if the outer symmetry group acts
transitively. Equivalently, one can take the following definition from~\cite{F1}.
\begin{Def} \label{defhomo}
A fermionic projector~$P$ is called {\bf{homogeneous}} if for
any~$x_0, x_1 \in M$ there is a permutation~$\sigma \::\: M \rightarrow M$
with~$\sigma(x_0)=x_1$ and an unitary transformation~$U$ such that
\[ P(\sigma(x), \sigma(y)) \;=\; U\: P(x,y)\: U^{-1}\spc \forall\, x,y \in M\:. \]
\end{Def}
We also refer to the permutation~$\sigma$ as a {\bf{translation}} in space-time,
which translates~$x_0$ to~$x_1$.
Since the local traces are not all the same, our minimizing fermion system is
certainly not homogeneous. Thus the homogeneity of discrete space-time is destroyed by the fermionic projector. Using a more graphic notion,
we also say that the translation symmetry of discrete space-time
is spontaneously broken by the fermionic projector.

Using the symmetry structure of the minimizing fermionic projector, we can
compute the lengths of the Bloch vectors analytically.
After a suitable rotation in~$\R^3$ we can assume that the
Bloch vectors of the equilateral triangle are proportional to the unit vectors
$$ \vec{e}_1 \;=\; \begin{pmatrix}1 \\ 0 \\ 0 \end{pmatrix},\qquad
\vec{e}_2 \;=\; \begin{pmatrix} -1/2 \\ \sqrt{3}/2 \\ 0 \end{pmatrix},\qquad
\vec{e}_3 \;=\; \begin{pmatrix} -1/2 \\ -\sqrt{3}/2 \\ 0 \end{pmatrix} , $$
whereas the two Bloch vectors on the straight line are positive multiples of the
unit vectors
$$ \vec{e}_4 \;=\; \begin{pmatrix} 0 \\ 0 \\ 1 \end{pmatrix} \qquad{\mbox{and}}\qquad
\vec{e}_5 \;=\; \begin{pmatrix}0\\0\\-1\\ \end{pmatrix}\:. $$
In view of the symmetries of the numerical minimizers, we take the ansatz
\[ \vec{v}_x \;=\; \alpha\, \vec{e}_x \quad {\mbox{for $x =1,2,3$}} \qquad {\mbox{and}} \qquad
\vec{v}_x \;=\; \beta\, \vec{e}_x \quad {\mbox{for $x =4,5$}} \]
with two parameters~$\alpha, \beta>0$. Imposing~(\ref{vrc5}), the relation~$\Tr(P)=2$ leads to
the condition~$3\alpha+2\beta=2$, leaving us with one free parameter~$\alpha \in (0, 2/3)$.
Using Proposition~\ref{prp41}, we can write the action as a function of $\alpha$:
\[ \Ss(\alpha)=10 \;\frac{1}{8}\;\alpha^4\,-\,18\;\alpha^3\,+\,15\;\alpha^2\,-\,6\;\alpha\,+\,1\,.\]
A straightforward calculation shows that there is precisely one real minimum at
$$ \alpha \;=\; -\frac{1}{9}\left(2+2\sqrt{17}\right)^{1/3}+\frac{4}{9}\left(2+2\sqrt{17}\right)^{-1/3}+\frac{4}{9}\;\approx\; 0.4077411555\:, $$
in agreement with our numerical results.
In particular, this calculation shows that the mi\-ni\-mum of the action is
indeed smaller than the action of the more symmetric configuration with~$\alpha = \beta$.

\subsection{Six to Nine Space-Time Points: Emergence of a Two-Dimensional Lattice}
We now discuss our numerical results for more than five space-time points.
For the local traces and the lengths of the Bloch vectors we always find
\[ \rho_x \;=\; |\vec{v}_x| \;=\; \frac{2}{m} \qquad \forall\, x \in M\:. \]
All the space-time points have time-like separation. But the minimizing fermionic projector
is not permutation symmetric.

\begin{figure}[t]
\begin{center}
 \includegraphics[width=5cm]{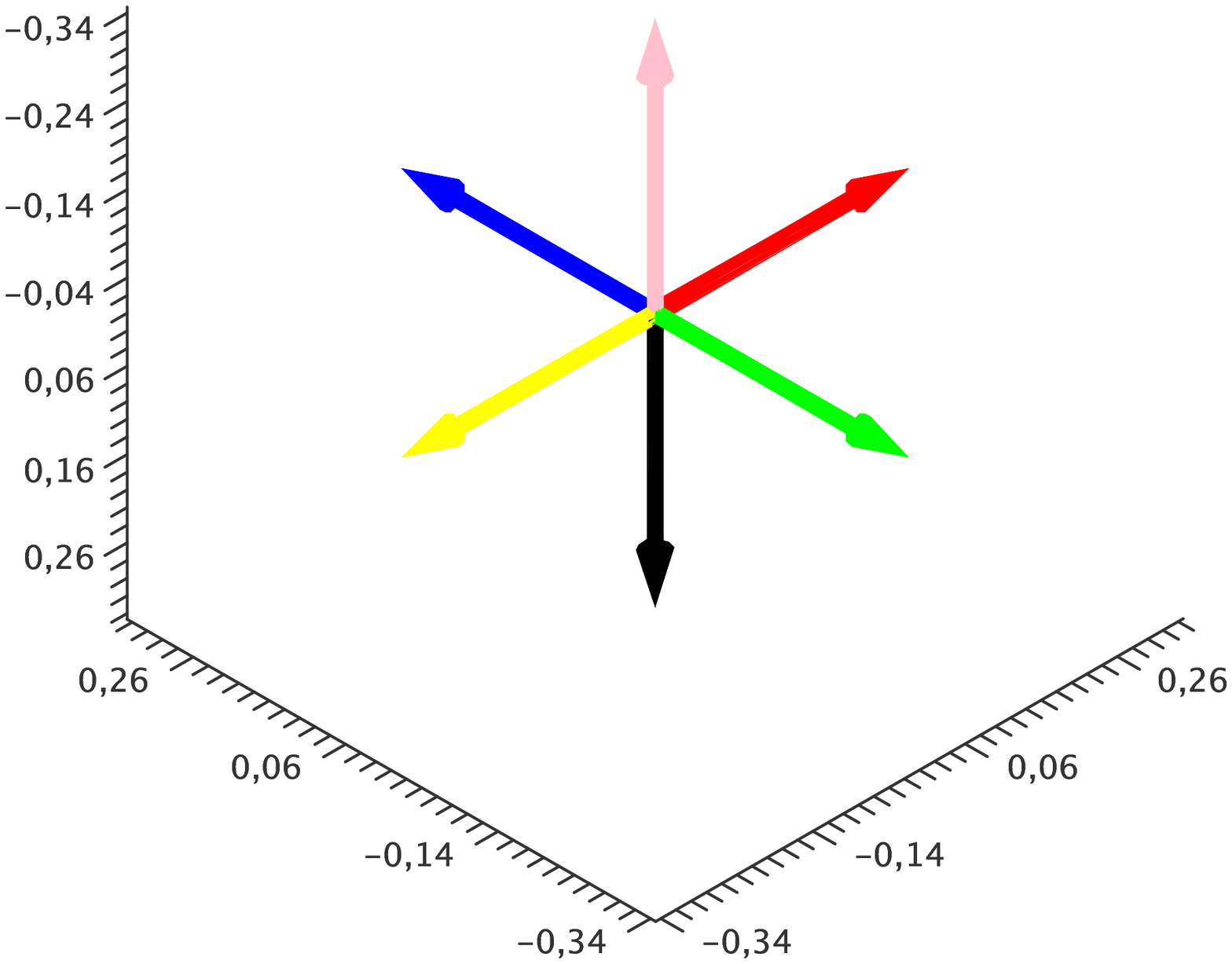}
 \includegraphics[width=5cm]{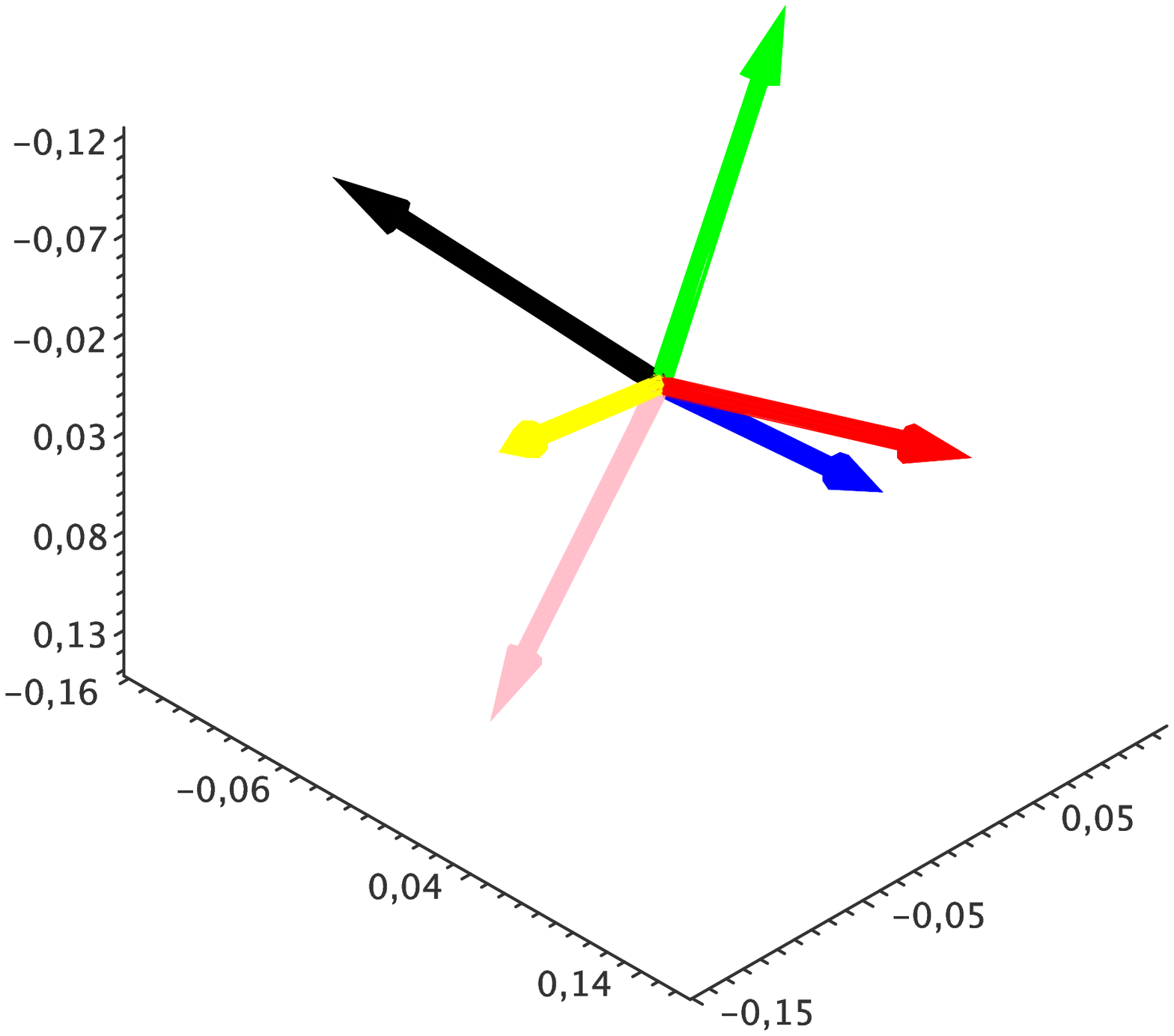}
 \caption{Bloch vectors for six space-time points, in an octahedral configuration (left)
 and a non-octahedral configuration (right)} \label{fig6}
\end{center}
\end{figure}

In the case of six space-time points, the most symmetric polyhedron with six vertices is the octahedron. The Bloch vectors of length $2/6$ which form an octahedron indeed
have the minimal action $\Ss=2/27$, and we found fermion systems which
realize these octahedral Bloch vectors.
However, we also found another type of minimizer which, within the accuracy of our numerics of nine digits,
has the same value of the action. For this minimizer, the angles between
the Bloch vectors are a bit different.
In Figure~\ref{fig6} the Bloch vectors of these two types of minimizers are shown.

\begin{figure}[t]
\begin{center}
 \includegraphics[width=4cm]{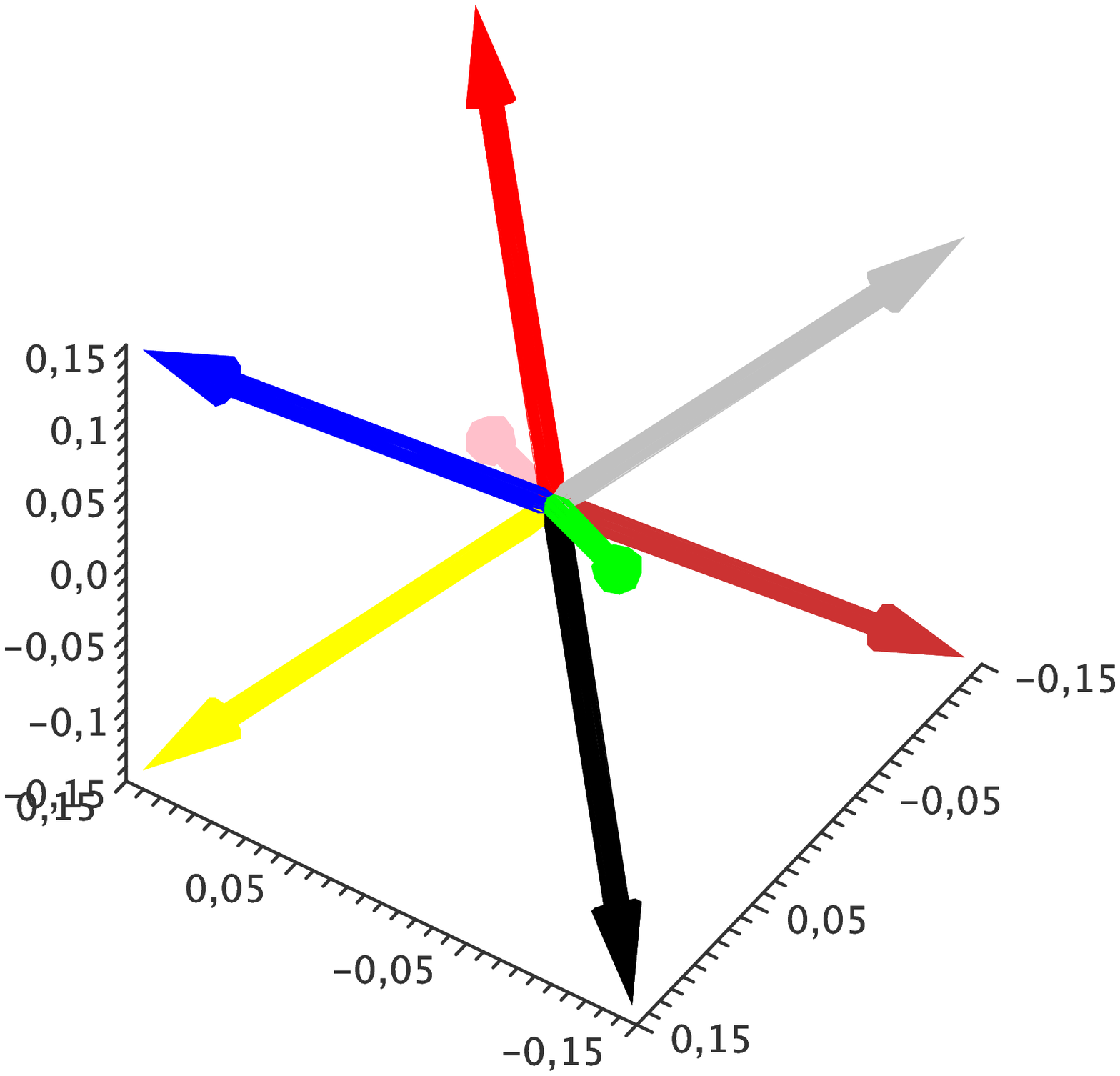}
 \includegraphics[width=4cm]{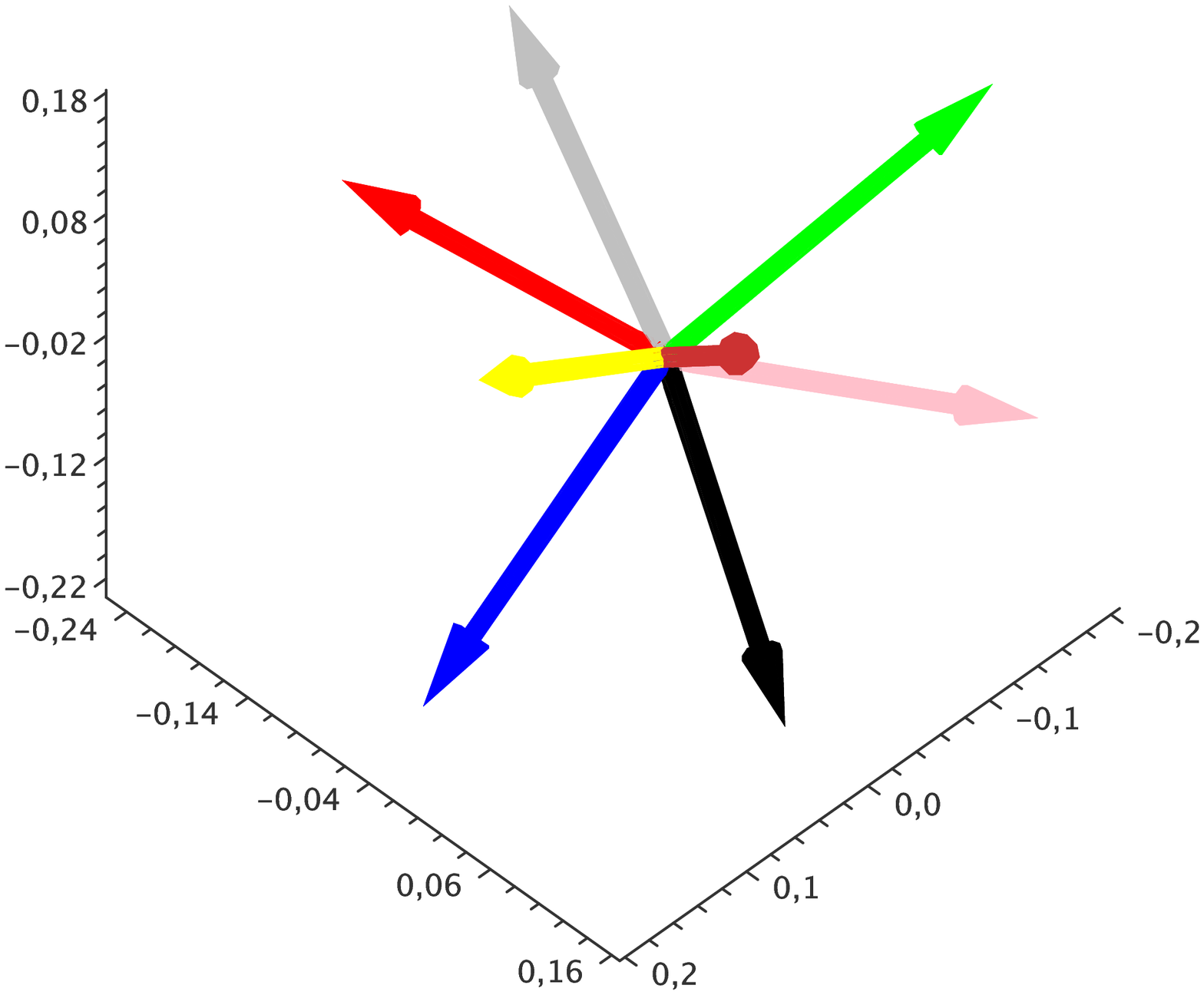}
 \includegraphics[width=4cm]{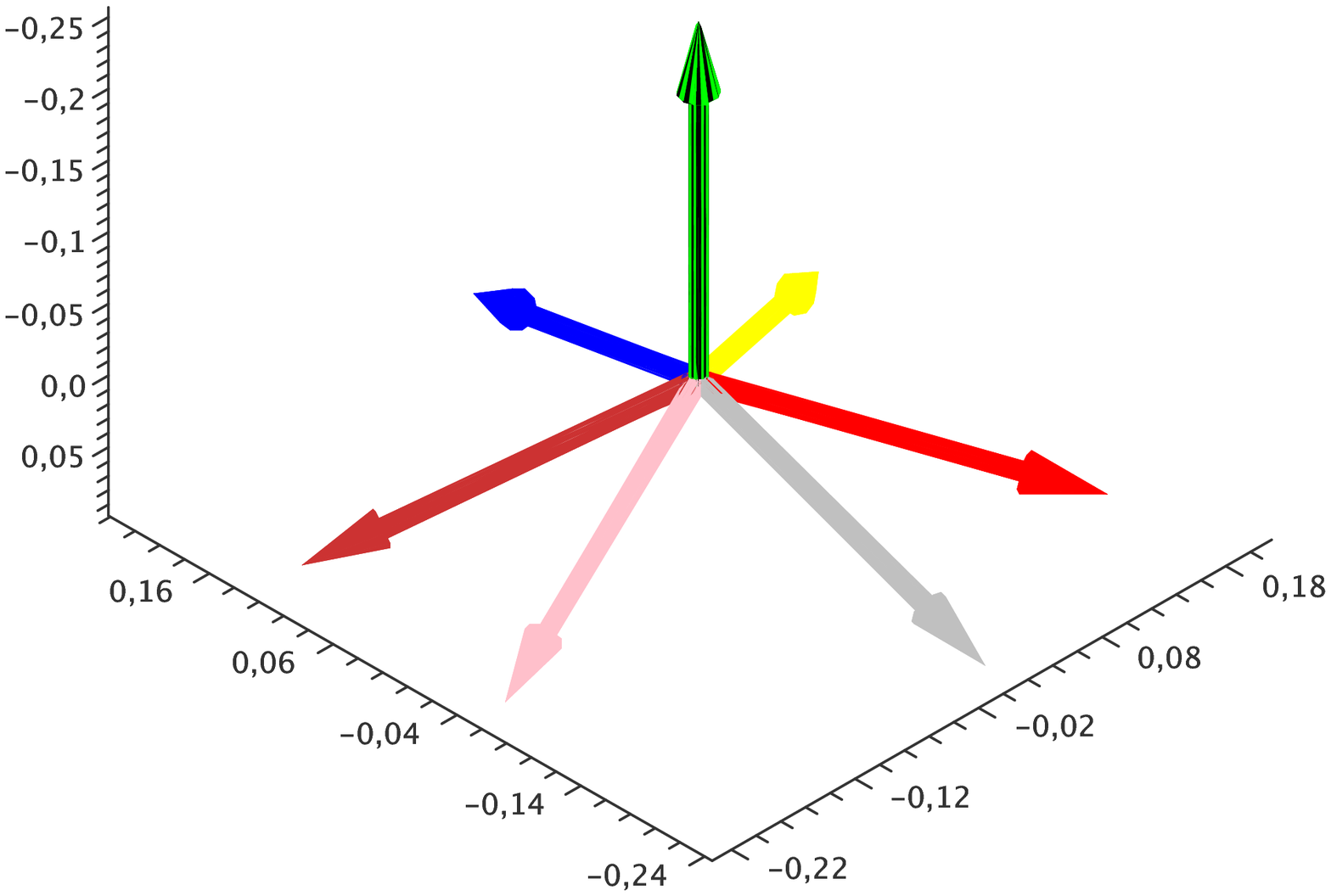}
\caption{Bloch vectors for eight space-time points, in a cubic configuration (left) and other configurations}
\label{fig8}
\end{center}
\end{figure}

Similarly, for eight space-time points the symmetric polyhedron with eight vertices is the cube.
The action of this configuration is $1/24$, and this is indeed the minimum.
But the minimizer is not unique. There are several local minima, as is illustrated in Figure~\ref{fig8}. Each minimizer achieves the same minimum up to nine valid digits.

\begin{figure}[t]
\begin{center}
 \includegraphics[width=5cm]{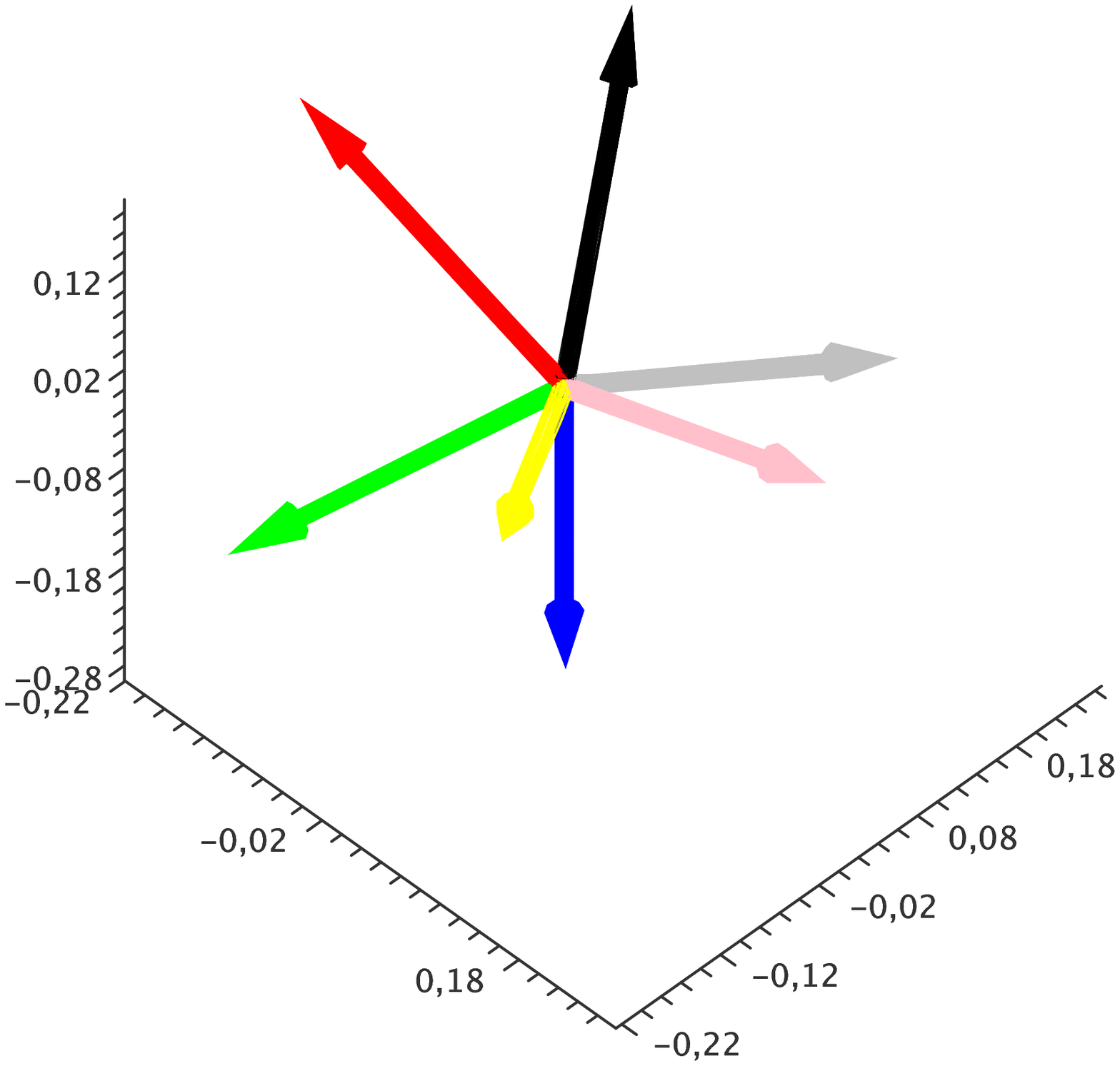}
 \includegraphics[width=5cm]{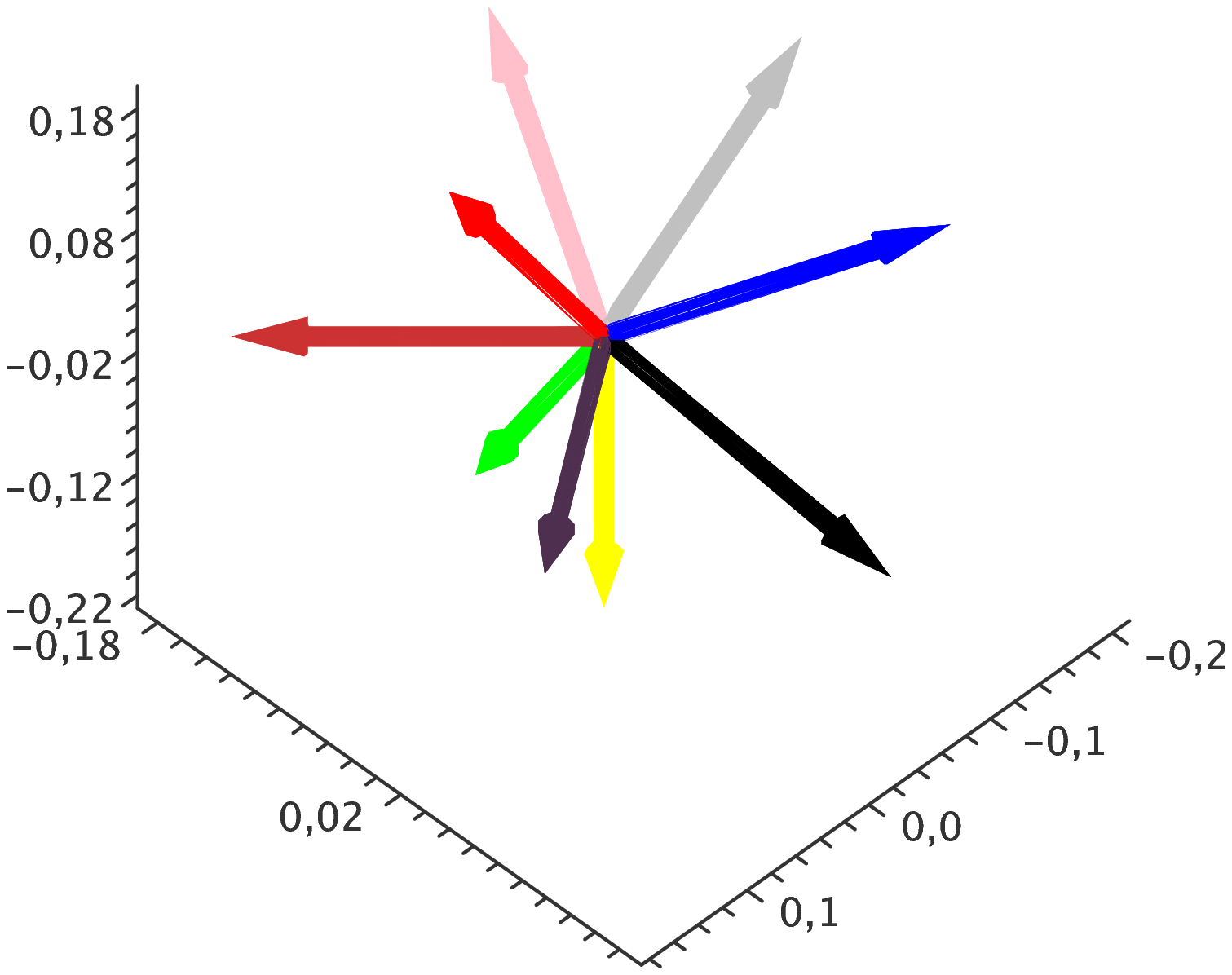}
\caption{Bloch vectors for seven (left) and nine (right) space-time points}
\label{fig7}
\end{center}
\end{figure}

For seven and nine space-time points, the Bloch vectors of typical minimizers are
shown in Figure \ref{fig7}.
These figures reveal the qualitative behavior of two-particle systems for many
space-time points: For large~$m$, the Bloch vectors of the minimizers
have at least approximately the same length~$2/m$ and can thus be identified with points on a
two-dimensional sphere of radius~$2/m$. The critical variational principle aims
at distributing these points uniformly on the sphere, i.e.\ it tries to maximize 
a certain distance function between these points.
For many space-time points, the resulting structure looks locally like a two-dimensional lattice
on a sphere. The qualitative behavior should be very similar to the situation described in
the survey article~\cite{SK}, where a general class of repulsive forces between points on the
sphere is considered.

\section{Two-Particle Systems for the Variational Principle with Constraint}
\setcounter{equation}{0}
\subsection{Two Space-Time Points}
As in Section~\ref{sec43}, we first discuss our numerical results and then construct
the minimizers analytically.
The first question is for what values of~$\kappa$ the constraint~(\ref{cfamily})
can be fulfilled. Our numerics (see Appendix~\ref{appB} for details) shows that
this is possible for all~$\kappa \geq 2$. For any~$\kappa$ in this range, we found
up to gauge transformations a unique minimizer of the variational principle~(\ref{varpr}).
This minimizer has permutation symmetry. If~$\kappa=2$, the
minimizing fermion system coincides precisely with that of the the critical case
as constructed in Section~\ref{sec42}. In the case~$\kappa>2$, the systems
are different because the length of the Bloch vectors is strictly larger than the
local trace.
All space-time points have timelike separation, and thus the discrete causal structure
is trivial.

The variational principle with constraint~(\ref{varpr}) could be treated analytically.
In order to keep the analysis reasonably simple, we only consider fermion systems
with permutation symmetry. Such systems were already constructed in~\cite{F1}.
\begin{Lemma} \label{lemma51} In the case~$f=2$ and~$m=2$, for any~$\vartheta \geq 0$, 
the fermion system corresponding to the fermion matrix
\beq \label{fm2}
\Psi \;=\; \left( \begin{matrix}  \sinh \vartheta & 0 \\
0 & \cosh \vartheta \\[.5em]
0 & \sinh \vartheta \\
\cosh \vartheta & 0
\end{matrix} \right)
\eeq
has the outer symmetry group~${\mathcal{O}}=S_2$.
Every fermionic projector with this outer symmetry group is gauge equivalent to
the fermionic projector corresponding to~(\ref{fm2}) for some~$\vartheta \geq 0$.
The local density matrices~(\ref{Fxdef}) are of the form~(\ref{Fxd}) with\begin{eqnarray}
\rho_1 &=& \rho_2 \;=\; 1 \label{52} \\
\vec{v}_1 &=& -\vec{v}_2 \;=\; \left(1 + 2 \sinh^2 \vartheta \right)
\left( \begin{matrix} 0 \\ 0 \\ -1 \end{matrix} \right) . \label{53}
\end{eqnarray}
\end{Lemma}
{\Proof} In~\cite[Example~3.2]{F1} it is shown that, after a suitable gauge transformation,
the fermionic projector can be written in the general form
\[ P \;=\; \left( \begin{matrix}
-\sinh^2 \alpha & 0 & 0 & \sinh \alpha \cosh \alpha \\
0 & \cosh^2 \beta & -\sinh \beta \cosh \beta & 0 \\
0 & \sinh \beta \cosh \beta & -\sinh^2 \beta & 0 \\
-\sinh \alpha \cosh \alpha & 0 & 0 & \cosh^2 \alpha \end{matrix} \right) \]
 with~$\alpha, \beta \in \R$, and by an additional gauge transformation we can arrange
 that~$\alpha, \beta \geq 0$. For a permutation symmetric system, the local traces~$\rho_1$
 and~$\rho_2$ must clearly coincide. Computing them, we obtain the
 condition~$\alpha=\beta$. This gives rise to the fermion matrix~(\ref{fm2}).
 The relations~(\ref{52}, \ref{53}) follow by a straightforward computation.
 \QED
 For permutation symmetric systems, we can analyze the variational principle
 with constraint in closed form.
\begin{Prp} Considering the variational principle~(\ref{varpr}) in the class of
fermionic projectors with the outer symmetry group~$\Oo=S_2$, the constraint can be fulfilled
if and only if~$\kappa \geq 2$. Then
\beq \label{minact}
\min \sum_{x,y=1}^2 |A_{xy}^2| \;=\;  \sqrt{\kappa-1} \:+\: \frac{\kappa}{2}\:,
\eeq
and the minimum is attained by the fermionic projector corresponding to the
fermion matrix~(\ref{fm2}) with
\beq \label{vval}
1+2 \sinh^2 \vartheta \;=\; (\kappa-1)^{\frac{1}{4}}\:.
\eeq
The corresponding Lagrange multiplier~$\mu$ in~(\ref{Lagdef}) has the value
\beq \label{muval}
\mu \;=\; \frac{1}{2} \left( 1 + \frac{1}{\sqrt{\kappa-1}} \right) .
\eeq
\end{Prp}
{\Proof} Setting~$v=|\vec{v}_1|=|\vec{v}_2| \geq 1$, a straightforward calculation using
Lemma~\ref{lemma51} and Proposition~\ref{prp41} yields
\begin{eqnarray*}
{\mbox{for $A_{11}$ and~$A_{22}$}}: &\qquad&\lambda_\pm=
\frac{1}{4} \left(1+v^2 \pm 2v \right) \;=\; \frac{1}{4} \left( 1 \pm v \right)^2 \\
{\mbox{for $A_{12}$ and~$A_{21}$}}: &&\lambda_+ =  \lambda_-=
\frac{1}{4}\: (1-v^2) \;=\;  \frac{1}{4} \:(1+v)(1-v)\:,
\end{eqnarray*}
and a short computation gives
\begin{eqnarray*}
\sum_{x,y=1}^2 |A_{xy}|^2 &=& 2 \left( |A_{11}|^2 + |A_{12}|^2 \right)
\;=\; 1+v^4 \\
\sum_{x,y=1}^2 |A_{xy}^2| &=& 2 \left( |A_{11}^2| + |A_{12}^2| \right)
\;=\; \frac{1}{2}\: (1+v^2)^2\:.
\end{eqnarray*}
Hence the constraint~(\ref{cfamily}) reduces to the equation
\[ v^4 \;=\; \kappa-1\:. \]
Hence the condition~$v \geq 1$ can be fulfilled only if~$\kappa \geq 2$, and in this case~$v=(\kappa-1)^{\frac{1}{4}}$.
This yields~(\ref{vval}) and~(\ref{minact}). To determine the Lagrange multiplier,
we first compute the action~(\ref{Smdef}),
\[ \Ss_\mu \;=\; v^2 + (1-\mu) \left(1+v^4 \right) \:, \]
and demand that its $v$-derivative should vanish at~$v=(\kappa-1)^{\frac{1}{4}}$.
\QED
It is worth noting that the value of the Lagrange multiplier~(\ref{muval}) is strictly larger
than the critical value~$\mu=\frac{1}{2}$. Hence the corresponding action~${\mathcal{S}}_\mu$,
(\ref{Smdef}, \ref{Lagdef}), is unbounded below. We thus have an example where a
minimizer of the variational principle with constraint~(\ref{varpr}) fails to be a minimizer
of the corresponding auxiliary variational principle~(\ref{auxvp}).

\subsection{Three Space-Time Points: Emergence of a Discrete Causal Structure}
For systems with three space-time points, we can study the emergence of a
non-trivial discrete structure in detail. We begin by a rigorous analysis of the variational
principle with constraint for permutation symmetric fermion systems. We will
show that for~$\kappa$ larger than a critical value, space-like separation appears.
Afterwards, we will discuss our numerical results, which confirm the analytic
picture, except that for very large values of~$\kappa$ in addition the permutation
symmetry is spontaneously broken.

\begin{figure}[t]
\begin{center}
 \includegraphics[width=16cm]{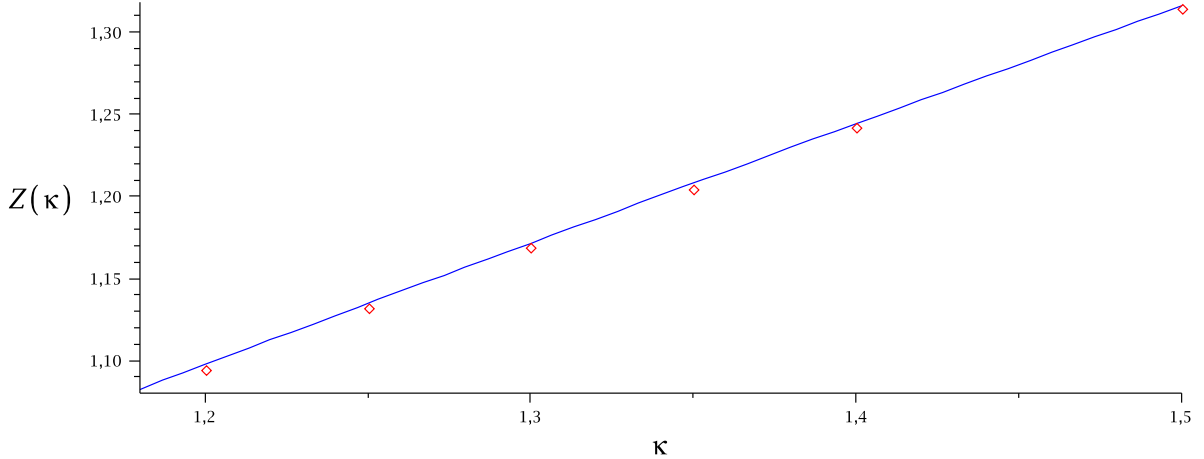}
\caption{Minimal value of $\Zz = \sum_{x,y\in M} |A_{xy}^2|$ as a function of~$\kappa$}
\label{m3plot}
\end{center}
\end{figure}

The permutation symmetric fermion systems were characterized in Lemma~\ref{lemma44},
making it possible to analyze the variational principle in detail.
\begin{Prp} \label{prp53}
Considering the variational principle~(\ref{varpr}) in the class of
fermionic projectors with the outer symmetry group~$\Oo=S_3$, the constraint (\ref{cfamily}) can be fulfilled
if and only if~$\kappa \geq \frac{2}{3}$. Then the local correlation matrices are of the form (\ref{Fxd}) with
\[ \rho_x\;=\;\frac{2}{3} \quad \mbox{and}\quad|\vec{v}_x|\;=\;\begin{cases}
\displaystyle \frac{1}{3}\:(72\kappa-32)^{\frac{1}{4}} &\displaystyle  \textit{if} \;\kappa\; \leq\; 
\frac{68}{81} \\[0.3cm]
\displaystyle \frac{1}{9}\:(12+6\sqrt{-32+81\kappa})^{\frac{1}{2}} & \displaystyle \textit{if} \;\kappa \;>\; \frac{68}{81}\:.
           \end{cases} \]
The function~(\ref{target}) takes the value
\beq\label{Agleichung}\
\Zz[P] \;=\;\begin{cases}
\displaystyle \frac{2}{9}\: (\sqrt{18\kappa-8}+1) &\displaystyle 
\textit{if} \;\kappa\; \leq\; \frac{68}{81} \\[0.3cm]
\displaystyle \frac{8}{81}\: (2+ \sqrt{-32+81\kappa})+\frac{\kappa}{2} &\displaystyle 
\textit{if} \;\kappa \;>\; \frac{68}{81}\:.
                             \end{cases} \eeq
In the case $\kappa \leq \frac{68}{81}$ all space-time points $(x,y)$ are timelike separated. If conversely $\kappa > \frac{68}{81}$, different space-time points are spacelike separated. 
\end{Prp}
{\Proof} In the proof of Proposition \ref{prp45} we already computed the roots of the closed chains (\ref{lampm1},\ref{lampm2}) of a fermionic projector with $S_3$-symmetry. The closed chains $A_{xx}$ always have the same spectral weight:
\[ |A_{xx}|\:=\:\frac{1}{18}\:(4+9v^2) \quad \mbox{for all} \quad x\in M. \]
Regarding different space-time points $x\neq y$ there are two cases to look at. If $v\leq \frac{4\sqrt{3}}{9}$ the closed chain $A_{xy}$ has two real roots and therefore 
\[ |A_{xy}|\:=\:\frac{1}{36}\:(8-9v^2) \quad \mbox{for all} \quad x\neq y. \]
The constraint~$P\in \Pp(\kappa)$ gives the condition
\[ 32+81v^4\:=\:72\kappa, \]
which is fulfilled exactly in the case $v=\frac{1}{3}(72\kappa-32)^{\frac{1}{4}}$. The condition $v\leq \frac{4\sqrt{3}}{9}$ is therefore equivalent to $\kappa \leq \frac{68}{81}$.\ \\
In the remaining case $v>\frac{4\sqrt{3}}{9}$ we obtain a pair of complex conjugate roots and therefore
\[ |A_{xy}|\:=\:2 \sqrt{\lambda_+ \lambda_-}\:=\:\frac{1}{18}(4-9v^2) \quad\mbox{for all}\quad x\neq y. \]
The constraint~$P\in \Pp(\kappa)$ gives the condition
\[ 243v^4-72v^2+48\:=\:108\kappa,\]which is fulfilled if and only if
$$v=\frac{1}{9}(12+6\sqrt{-32+81\kappa})^{\frac{1}{2}}.$$
Writing~$\Zz$ as
$$ \Zz \;=\;\Ss+\frac{\kappa}{2}$$
and using (\ref{actionm3}) gives (\ref{Agleichung}). 
{\QED}
Comparing the values of $\Zz$ of the above fermionic projectors to the minimal values calculated numerically, one finds that the fermionic projector with full $S_3$-symmetry is the minimizer
for small~$\kappa$. If however~$\kappa \geq 1$, we found fermionic projectors numerically
which realize even smaller values of~$\Zz$.
This is shown in Figure \ref{m3plot}, where the curve corresponds to the analytic minimizers with $S_3$-symmetry, whereas the single points are the numerical minimizers. Since
these points lie below the curve, the corresponding fermionic projectors cannot
have~$S_3$-symmetry. Hence for the minimizers for large~$\kappa$, the
permutation symmetry is spontaneously broken.

\begin{figure}
\begin{center}
 \includegraphics[width=13cm]{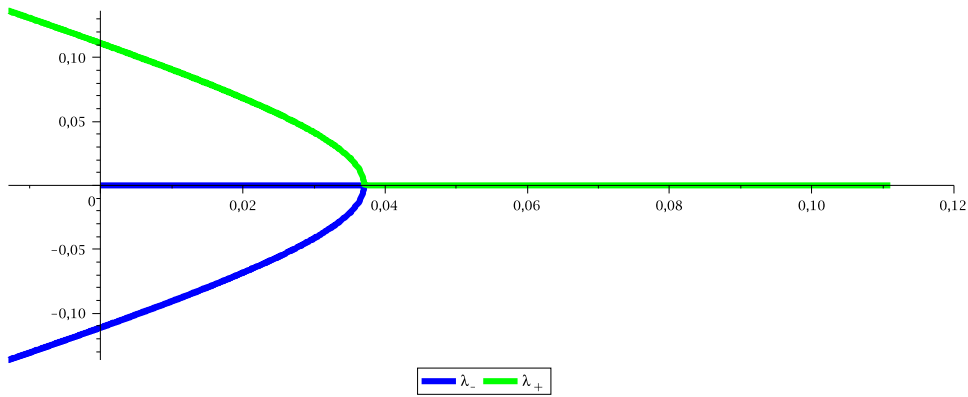}
 \caption{Plots of $\lambda_+$ and $\lambda_-$ for varying~$\kappa$}
\label{m3lambda}
\end{center}
\end{figure}
In the case~$\kappa > \frac{68}{81}$, the discrete causal structure
is also a {\em{causal set}} (see for example~\cite{csets}), albeit in a trivial way where
each point has timelike separation only from itself.
We point out that the discrete causal structure of Definition~\ref{dcs} differs from
the definition of a causal set in that it does not distinguish
between future and past directed separations. Thus in order to obtain a causal set,
one would have to order the points in each time-like separated pair in such a way that the
resulting partial ordering is transitive. It is an open question whether this can be done
for general fermion systems in discrete space-time.

\appendix
\section{Numerical Implementation of the Critical Case}
\setcounter{equation}{0} \label{appA}
We first bring the critical Lagrangian~(\ref{Lsimp}) into the form which is most convenient
for the numerical analysis. The roots of the characteristic polynomial of a $2\times 2$-matrix $A_{xy}$ are
\[ \lambda_\pm=\frac{1}{2}\left( \Tr(A_{xy})\pm \sqrt{\Delta_{xy}} \right) , \]
where
\[ \Delta_{xy}=\Tr (A_{xy})^2-4\det A_{xy}.\]
If~$\Delta_{xy}<0$, the $\lambda_\pm$ form a complex conjugate pair, and the 
Lagrangian vanishes. If conversely~$\Delta_{xy} \geq 0$, $\lambda_+$ and~$\lambda_-$
are both real and according to~(\ref{ssign}) have the same sign. Thus in this case
the critical Lagrangian~(\ref{Lsimp}) simplifies to~$\Ll = (\lambda_+-\lambda_-)^2/2
= \Delta_{xy}/2$. We conclude that
\[ \Ll[A_{xy}] \;=\; \frac{1}{2}\: \Delta_{xy}\: \Theta(\Delta_{xy})\:, \]
where~$\Theta$ is the Heaviside function, and thus
\[ \Ss \;=\; \frac{1}{2}\: \sum_{x,y\in M} \Delta_{xy}\: \Theta(\Delta_{xy})\:. \]
We are now making use of the gauge freedom. For every non-vanishing $2\times 2$-matrix there exists a
unitary~$U\in U(1,1)$ which transforms the matrix in exactly one of the following forms:
\[ \begin{pmatrix}
	0 & v \\
	u & w
	\end{pmatrix}
	\quad \mbox{ or } \quad
	\begin{pmatrix}
	u & v \\
	0 & w
	\end{pmatrix}
	\quad \mbox{ or } \quad
	\begin{pmatrix}
	u & w_1 \\
	u & w_2
	\end{pmatrix}, \]
where $u,v \geq 0$ and $w,w_1,w_2 \in \C$. We now let~$\Psi=(\psi_1,\psi_2)$ be a fermion matrix for two particles, so 
\begin{equation}\label{bed}
	\bra \psi_i\:|\:\psi_j\ket \;=\;- \delta_{ij}\:.
\end{equation}
We can assume that after a suitable gauge transformation, every local fermion matrix $E_x\Psi$ is of the first form:
\beq \label{locferm}
E_x\Psi \;=\; \begin{pmatrix} 0 & v_x \\ u_x & w_x  \end{pmatrix},
\eeq
where $u_x,v_x \geq 0$, and $ w_x=x_x+iy_x\in \mathbb{C}$.
This is justified as follows. For two space-time points, the explicit calculation 
in Section~\ref{sec42} shows that the minimizing fermion matrix is of the form (\ref{locferm}). For three space-time points, we studied all possible cases systematically
and found that the ansatz~(\ref{locferm}) gives rise to all minimizers. Finally, for more than three space-time
points we also did numerics with an alternative algorithm which does not use the gauge freedom. This algorithm, which is less efficient,  always gave the same minimizers as the ansatz~(\ref{locferm}).

A direct calculation using (\ref{locferm}) and  (\ref{PsiP}) gives the fermionic projector and the closed chains. Then the Lagrangians $\Ll[A_{xy}]$ map the vector
\[ \xi \;=\; (u_1,\ldots,u_m,v_1,\ldots,v_m,x_1,\ldots,x_m,y_1,\ldots,y_m) \]
to a real positive number, and summing over all space-time points gives the action $\Ss$. $\mathcal{S}$ is the function to be minimized, where the vector $\xi$ has to lie in the feasible set given by (\ref{bed}). This leads to four constraint functions mapping $\R^{4m}$ to $\R$. Namely, the normalization constraints are realized by the functions
\[ r_1(\xi)=\sum_{i=1}^m u_i^2 -1 \qquad \mbox{ and } \qquad r_2(\xi)=\sum_{i=1}^m (x_i^2+y_i^2-v_i^2)-1 \:, \]
whereas the orthogonality condition gives  rise to the two functions
\[ r_3(\xi)=\sum_{i=1}^m u_i x_i \qquad \mbox{ and }\qquad r_4(\xi)=\sum_{i=1}^m u_i y_i\:. \]
Our optimization problem is to 
\[ \mbox{ minimize } \mathcal{S}(\xi) \mbox{ where } r_i(\xi)=0 \mbox{ for } i=1,\ldots,4\:. \]
We solve this nonlinear optimization problem with constraints with the penalty-method (see \cite{opt}, Chapter~17), where each minimization step is carried out with the algorithm of Fletcher-Reeves (see \cite{opt}, Chapter~5). Although the critical action is only piecewise~$C^2$,
working with the gradient of~${\mathcal{S}}$ works well, probably because the
action is smooth near the minimizers. We choose the quadratic penalty-function $Q$ to be
\[ Q(\xi;L)=\Ss(\xi)+L \sum_{i=1}^4 r_i^2(\xi)\:. \]
More precisely, our algorithm proceeds as follows. Starting with a random vector $\xi_0^s$, the penalty parameter $L_0=1000$ and the tolerance $\tau_0=10^{-6}$, we minimize in each step $Q(.;L_k)$ for a fixed parameter $L_k$. For this minimization we use Fletcher-Reeves, starting at $\xi_k^s$, and terminate the algorithm if $\| \nabla Q(\xi_k;L_k)\|^2<\tau_k$ or if the loop is ran too often, i.e.\ 100000 times. If the calculated minimizer $\xi_k$ is feasible, i.e.\ if $\sum_{i=1}^4 r_i^2(\xi_k)>10^{-20}$, the algorithm stops. Else the penalty parameter is increased to $L_{k+1}=1.1\,L_k$ and the tolerance is reduced to $\tau_{k+1}=0.9\,\tau_k$. The calculated minimizer $\xi_{k+1}$ becomes the starting vector in the next step, $\xi_{k+1}^s=\xi_{k}$. The C-program leads to the following minimal actions depending on the number of space-time points $m$:
\begin{center}
\begin{tabular}{l||l|l|l|l|l}
m & 1 & 2 & 3 & 4 & 5 \\
\hline
$\mathcal{S}$ & 0 & 1 & 1/3 & 1/6 & 0.10701459\ldots \\
\hline \hline
m & 6 & 7 & 8 & 9 & 10 \\
\hline
$\mathcal{S}$ & 2/27 & 0.05442177\ldots & 1/24 & 0.0329218\ldots & 2/75
\end{tabular}
\end{center}

\section{Numerical Implementation of the Variational Principle with Constraint}
\setcounter{equation}{0} \label{appB}
We now describe the numerical implementation of the variational principle with
constraint~(\ref{varpr}). A major difference compared to Appendix~\ref{appA} is that
we do not use a gradient method. Namely, gradient methods did not work, because the
function~$\Zz$ does not seem to be smooth near the minimizers in the
class~${\mathcal{P}}(\kappa)$.

The smallest~$\kappa$ for which the set ${\mathcal{P}}(\kappa)$ is non-empty,
denoted by~$\kappa_{\rm{min}}$, is computed with the following algorithm.
\begin{description}
\item[Step 1:] choose all entries of an initial fermion matrix, i.e.\ the vectors $u_{1},\ldots,u_{f}$, randomly in~$\C^{2m}$, where the absolute values of all real and imaginary parts are smaller than $1$; introduce an initial distance $\delta$, set $\delta = 1$
\item[Step 2:] calculate the fermionic projector $P$ and for all $x,y \in M$ the closed chains $A_{xy}$
\item[Step 3:] calculate
$$\widetilde{\kappa}_{\rm{min}} = \sum_{x,y \in M} \left| A_{xy} \right|^2 + \sigma(u_{1},\ldots,u_{f})$$
with the penalty function
$$\sigma(u_{1},\ldots,u_{f})= L_{\rm{norm}} \sum_{i=1}^f \left| \bra u_{i} | u_{i} \ket + 1 \right| + L_{\rm{orth}} \sum_{i=1}^{f-1} \sum_{j=i+1}^f \left| \bra u_{i} | u_{j} \ket \right| \text{,}$$
where $L_{\rm{norm}}$ and $L_{\rm{orth}}$ are positive parameters which can be increased stepwise while running the algorithm; for the minimizer of $\widetilde{\kappa}_{\rm{min}}$ the value of the penalty function is zero (i.e.\ all normalization conditions are fulfilled),
and we get $\kappa_{\rm{min}}$
\item[Step 4:] in order to get neighboring fermion matrices choose increment vectors $\Delta u_{1}$, $\ldots$, $\Delta u_{f}$, multiply them by $\delta$; the performance of the algorithm increases with the number of considered neighboring fermion matrices; this number is to be chosen depending on the available CPU power and the size of the considered fermion system.
\item[Step 5:] calculate $\widetilde{\kappa}_{\rm{min}}$ for $u_{1} + \Delta u_{1}$, $\ldots$, $u_{f} + \Delta u_{f}$, there are two cases:
\begin{description}
\item[Case 1:] If there is a fermion matrix with a smaller $\widetilde{\kappa}_{\rm{min}}$, then consider it as the new initial matrix and go back to step 4.
\item[Case 2:] If none of the considered neighboring fermion matrices has a smaller $\widetilde{\kappa}_{\rm{min}}$, go to step 4 with a smaller $\delta$, for example put $\delta \longrightarrow \frac{3}{4} \delta$.
\end{description}
\item[STOP] when $\delta < 10^{-6}$
\end{description}

\begin{figure}[t]
\begin{center}
\includegraphics{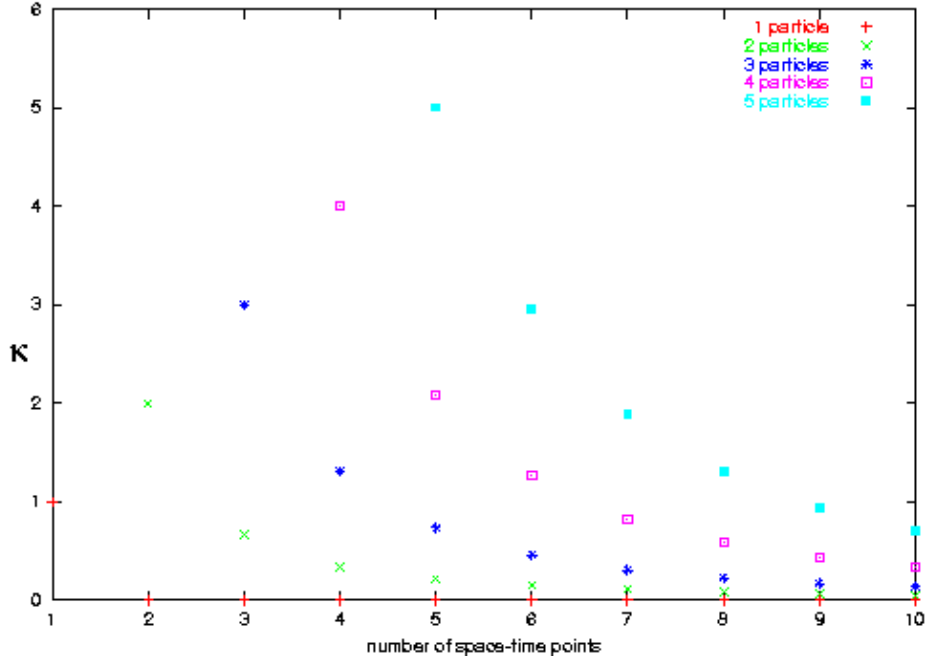}
\caption{Minimal value of $\kappa$ for various fermion systems}
\label{kappa_min}
\end{center}
\end{figure}

We considered fermion systems consisting of one to five particles and up to ten space-time points. The numerical results are shown in Figure \ref{kappa_min}. For one- and two-particle systems we find that the $\kappa_{\rm{min}}$ coincide precisely with the minima of the critical variational principle. Furthermore, for all considered fermion systems we observe that $\mathcal{S}_{\rm{min}} (\kappa_{\rm{min}})= \kappa_{\rm{min}}$. 

After having calculated $\kappa_{\rm{min}}$ for a given fermion system, we now solve the variational principle for any given~$\kappa \geq \kappa_{\rm{min}}$. We use the same algorithm with one exception: instead of minimizing $\widetilde{\kappa}_{\rm{min}}$, we minimize
$$\widetilde{\Zz}(\kappa) = \sum_{x,y \in M} \left| A_{xy}^{2} \right| + L_{\rm{side}} \left| \sum_{x,y \in M} \left| A_{xy} \right|^2 - \kappa \right| + \sigma(u_{1},\ldots,u_{f}) \text{,}$$
where $L_{\rm{side}}$ is another positive parameter. The additional penalty term $\left| \sum_{x,y \in M} \left| A_{xy} \right|^2 - \kappa \right|$ gives rise to the additional constraint equation $\sum_{x,y \in M} \left| A_{xy} \right|^2 = \kappa$.

We finally mention some numerical results which are not referred to from the main sections.
A disadvantage of the above algorithm is that the normalization conditions are difficult
to fulfill, and thus we had to discard many fermion configurations (for two particles and three space-time points, for example, we worked with 10 million neighboring matrices).
Therefore, it is worth considering whether the normalization conditions can be relaxed
without changing the minimizers.
To study this question, we introduced a new variational principle with a more general operator~$P$. This new variational principle is solved numerically, and we compare the results to the original variational principle with constraint.

\begin{Def}
A self-adjoint operator $A$ in an indefinite inner product space is called \textbf{positive} if $\bra u | Au \ket \geq 0$ for all $u \in H \text{.}$
\end{Def}

For a fermionic projector we have $\operatorname{Tr}(P)=f$, the rank of $P$ is equal to~$f$,
and $(-P)$ is positive due to the relation
$$
\bra u | (-P)u \ket = - \bra u | P^{2}u \ket = - \bra Pu | Pu \ket \geq 0
$$
Therefore, the next definition extends the class of fermionic projectors.

\begin{figure}
\centering
\includegraphics[height=8cm]{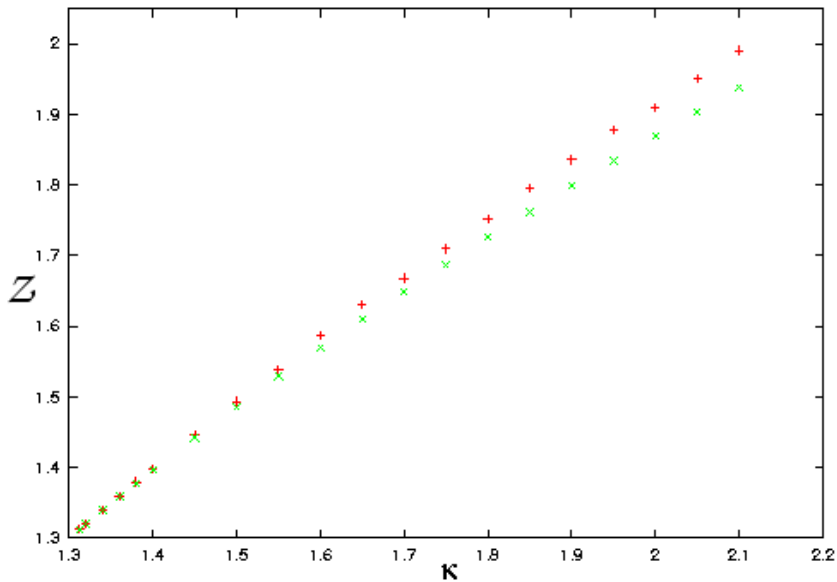}
\caption{Minimal~$\Zz$ for variations in~${\mathcal{P}}^f(\kappa)$ ($\times$) and in~${\mathcal{P}}(\kappa)$ ($+$) for the system with three particles and four space-time points} \label{fig34}
\end{figure}

\begin{Def}
An operator $P$ in an inner product space $(H,\bra.|.\ket)$ is \textbf{of the class $P^{f}$}, if $(-P)$ is positive, $\operatorname{Tr}(P)=f$ and $\operatorname{rk}(P) \leq f$.
\end{Def}
For a given parameter $\kappa > 0$ we define a family of operators
$$
\mathcal{P}^{f}(\kappa) = \left\{ P\ \text{of the class $P^{f}$ with} \sum_{x,y \in M} \left| A_{xy} \right|^2=\kappa \right\}
$$
We now introduce the modified variational principle
$$\text{minimize}\ \Zz \;=\; \sum_{x,y \in M} \left| A_{xy}^2 \right|\ \text{by varying}\ P\ \text{in
the class}\ \mathcal{P}^{f}(\kappa) \text{,}$$
where the number of particles $f$ and the discrete space-time $(H ,\bra . | . \ket,(E_{x})_{x \in M})$ are kept fixed. For this variational principle, the existence of minimizers is proved
in~\cite[Theorem~2.8]{F1}.

In order to solve this variational principal numerically, the existing algorithm must be modified.
In step 1 for $i \in \lbrace 1, \ldots , f \rbrace$ any vectors $u_{i}$ with $\bra u_{i} | u_{i} \ket < 0$ are admissible. This ensures the positivity of $(-P)$. The condition $\operatorname{Tr}(P)=f$ is realized by a simple rescaling of the vectors $u_{i}$. The penalty function is no longer needed, so here we choose $\sigma(u_{1},\ldots,u_{f})=0$, which reduces the number of numerical operations and makes the algorithm faster.

\begin{figure}[tbc]
\centering
\includegraphics[height=8cm]{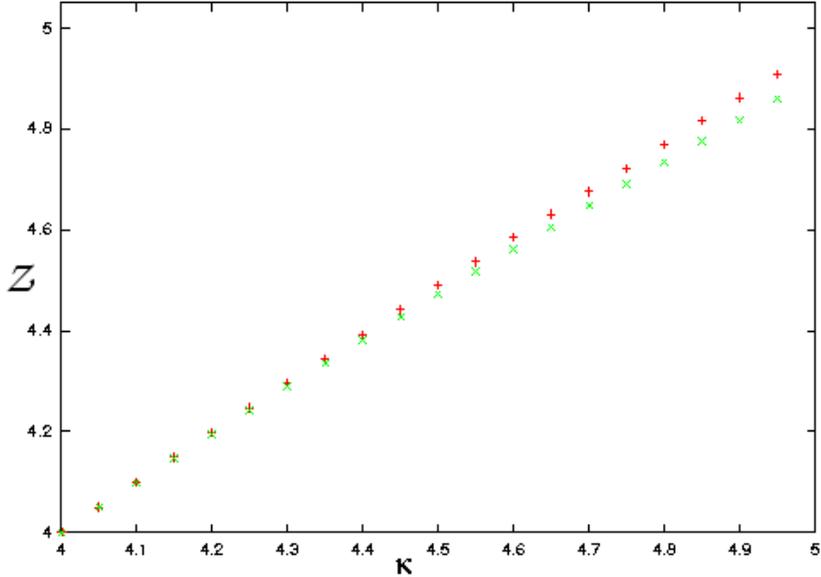}
\caption{Minimal~$\Zz$ for variations in~${\mathcal{P}}^f(\kappa)$ ($\times$) and in~${\mathcal{P}}(\kappa)$ ($+$) for the system with four particles and four space-time points} \label{fig44}
\end{figure}

Again we first calculate $\kappa_{\rm{min}}$ which is the smallest possible $\kappa$. We find for all considered fermion systems that the value of~$\kappa_{\rm{min}}$ coincides
with that for the original variation principle. In Figure~\ref{fig34} and Figure~\ref{fig44}
we compare the minimal values of~$\Zz$ for operators in the class~${\mathcal{P}}^f$
(points~$\times$ in green) with the corresponding values for projectors (points~$+$ in red).
In general, the minimum in the class~${\mathcal{P}}^f$ is strictly smaller, which was
expected because we minimize over a larger class of operators.
However, it is remarkable that for~$\kappa=\kappa_{\rm{min}}$ the minima
are the same. It turns out that the minimizers in the class~${\mathcal{P}}^f$
are actually projectors. Hence in this case, the minimizers in~${\mathcal{P}}^f$
automatically satisfy the normalization conditions. \\

\noindent
{\em{Acknowledgments:}} We thank Joel Smoller for helpful
comments on the manuscript.

\noindent
NWF I -- Mathematik,
Universit{\"a}t Regensburg, 93040 Regensburg, Germany, \\
{\tt{Alexander.Diethert@googlemail.com}}, \\
{\tt{Felix.Finster@mathematik.uni-regensburg.de}}, \\
{\tt{Daniela.Schiefeneder@mathematik.uni-regensburg.de}}

\end{document}